\documentclass[twocolumn,english,aps,prx,floatfix,amssymb,superscriptaddress,longbibliography]{revtex4}
\usepackage[latin9]{inputenc}
\usepackage{verbatim}
\usepackage{float}
\usepackage{amsmath}
\usepackage{amssymb}
\usepackage{graphicx}
\usepackage{color}
\usepackage{xcolor}
\usepackage{tikz}
\usepackage[normalem]{ulem}
\usepackage{bbold}
\DeclareMathOperator{\sgn}{sgn}

\usepackage[bookmarks=false,linkcolor=blue,urlcolor=blue,colorlinks,citecolor=blue]{hyperref}
\makeatletter

\newcommand{\be}{\begin{equation}}
\newcommand{\ee}{\end{equation}}
\newcommand{\bea}{\begin{eqnarray}}
\newcommand{\eea}{\end{eqnarray}}

\begin{document}

\title{Boundary Quantum Phase Transitions in the Spin $\frac{1}{2}$ Heisenberg Chain with Boundary Magnetic Fields}
\author{Parameshwar R. Pasnoori}
\thanks{These two authors contributed equally.}
\affiliation{Department of Physics,
University of Maryland, College Park, Maryland 20742 USA}
\affiliation{ Laboratory for Physical Sciences, 8050 Greenmead Dr, College Park, Maryland 20740 USA}
\author{Junhyun Lee}
\thanks{These two authors contributed equally.}
\affiliation{Department of Physics and Astronomy, Center for Materials Theory, Rutgers University, Piscataway, NJ 08854-8019 USA}
\author{J. H. Pixley}
\affiliation{Department of Physics and Astronomy, Center for Materials Theory, Rutgers University, Piscataway, NJ 08854-8019 USA}
\affiliation{Center for Computational Quantum Physics, Flatiron Institute, New York, New York 10010, USA}
\author{Natan Andrei}
\affiliation{Department of Physics and Astronomy, Center for Materials Theory, Rutgers University, Piscataway, NJ 08854-8019 USA}
\author{Patrick Azaria}
\affiliation{Laboratoire de Physique Th\'eorique de la Mati\`ere Condens\'ee, Sorbonne Universit\'e and CNRS, 4 Place Jussieu, 75252 Paris, France}
\email{pparmesh@umd.edu}
\email{junhyun.lee@physics.rutgers.edu}

\begin{abstract}
We consider the 
spin $\frac{1}{2}$ Heisenberg chain with boundary magnetic fields and analyze it using a combination of Bethe ansatz and density matrix renormalization group (DMRG) techniques. We show that the system exhibits several different ground states which depend on the orientation of the boundary magnetic fields. When both the boundary fields take equal values greater than a critical field strength, each edge in the ground state accumulates a fractional spin which saturates to spin $\frac{1}{4}$, which is similar to systems exhibiting symmetry protected topological phases (SPT). Unlike in SPT systems, the fractional boundary spin in the Heisenberg 
spin chain is not a genuine quantum number since the variance of the associated operator does not vanish, this is due to the absence of a bulk gap.
The system  
exhibits high energy bound states when the boundary fields take values greater than the critical field. All the excitations in 
the system can be sorted out into towers whose number 
depends on the number of bound states exhibited by the system. 
As the boundary fields are varied, in addition to the ground state phase transition, we find that the system may undergo
an eigenstate phase transition (EPT) where the number of towers of the Hilbert space changes.  We further inquire how the EPT reflects itself on local ground state properties by computing the magnetization profile $\langle S^z_j \rangle$ using DMRG.
We identify a clear qualitative change from low edge fields to  high edge fields when crossing the critical field. We though are unable to conclude on the basis of our data that EPT corresponds to a genuine phase transition in the ground state. 

\end{abstract}
\maketitle
\section{Introduction}


The Heisenberg model is  one of the most celebrated models in condensed matter and 
statistical physics.  It lies at the cornerstone of our understanding of many  physical phenomenon which, besides magnetism, 
consists of integrability \cite{Bethe1931,Sklyannin,YangYang1}, many body localization \cite{MBL1,MBL2} 
and
out of equilibrium dynamics \cite{dynamics1,dynamics2,dynamics3}. 
Thanks to available analytic and numerical methods the model is quite well understood
in one spatial dimension.
This is 
partly due
to the fact that the one dimensional spin $S=1/2$ Heisenberg model, which is also known as the XXX spin chain, is integrable.
At the same time,
the model  can be probed experimentaly either 
in solid-state compounds comprising quasi one-dimensional spin chains in KCuF$_3$ \cite{Steiner,Satija,Nagler,Tennant,Maillet} or 
more recently in ultra-cold atom realizations of the spinful Bose-Hubbard model \cite{PhysRevX.11.041054,jepsen2022long}. 
Since it was first solved by Bethe \cite{Bethe1931}, the spin chain with periodic boundary conditions has been very well studied. Both  the ground state and  the low energy excitations properties are well understood \cite{YangYang1,YangYang2,YangYang3,BABELON,Boos}. The system is non magnetic and supports massless  spin $\frac{1}{2}$ excitations named spinons. Besides this, integral representations of correlation functions have been obtained \cite{Kitanine,Shiroishi,Takahashi}. Spin chains with open boundaries have also been intensely studied after the Yang-Baxter algebra was generalized to systems with open boundaries by  Sklyannin and Cherednik \cite{Sklyannin, Cherednik}. For the   XXX spin chain with open boundaries the ground state, bulk excitations and physical boundary $S-$matrices have been found \cite{GrisaruMezincescNepomechie}. More generally the effects of boundary fields \cite{Cao02,Cao21,Sun_2019} have also been investigated and spin chains with non diagonal boundary fields have been solved \cite{ODBA}. 

In this work we shall be interested in the XXX spin chain with magnetic fields at its edges. Although the subject has been studied to some extent, for a quantum impurity \cite{XXXkondo} as well as for a classical one 
(e.g.  with boundary magnetic fields) we find that some issues remain to be clarified and explored in the light of new developments related with one dimensional topological phases and eigenstate phase transitions. To the best of our knowledge the results that will be presented in this work have not been found before. 

The Hamiltonian of the XXX spin chain with boundary magnetic fields is given by
\bea
H=\sum_{j=1}^{N-1}\sum_{\alpha=x,y,z}\sigma^{\alpha}_j\cdot \sigma^{\alpha}_{j+1} +h_L\; \sigma^z_1+h_R\; \sigma^z_N, \label{eq:H}\eea
 where $\sigma^{\alpha}_j$ are the Pauli matrices acting on the spin space at site $j$,  $h_L$ and $h_R$ are  boundary magnetic fields acting at sites $j=1$ and $j=N$ respectively, and $N$ is the number of sites.  The boundary magnetic fields break the  $SU(2)$ spin  symmetry down to the $U(1)$ group of rotation around the ``$z$'' axis. There exists no other symmetries except when $h_L=h_R$, where the model displays space parity invariance $\mathbb{P}$. Despite this, the system possesses a useful isometry obtained by simultaneously flipping all spins as well as reversing the orientation of the boundary fields:
 \bea \label{z2}H \rightarrow  \prod_{i=1}^{N}\sigma^{x}_i H \sigma^{x}_i, \;\;h_L\rightarrow-h_L, \;\; h_R\rightarrow-h_R.\eea
 The latter isometry is a symmetry of the phase diagram of the model in the plane $(h_L,h_R)$.
The Hamiltonian (\ref{eq:H}) is integrable by the method of the Algebraic Bethe Ansatz. The related Bethe equations has been first obtained in Ref. \cite{Sklyannin,GrisaruMezincescNepomechie,skorik,kapustinxxz} where a preliminary analysis was given. In the present work we shall extend their analysis and present a more thorough picture of the phase diagram associated with (\ref{eq:H}), thus  providing important  new results that have not been, to the best of our knowledge, present in the literature. But before going into more details   let us first discuss qualitatively our main results. 

\begin{center}
\begin{figure}[!h]
\includegraphics[width=1\columnwidth]{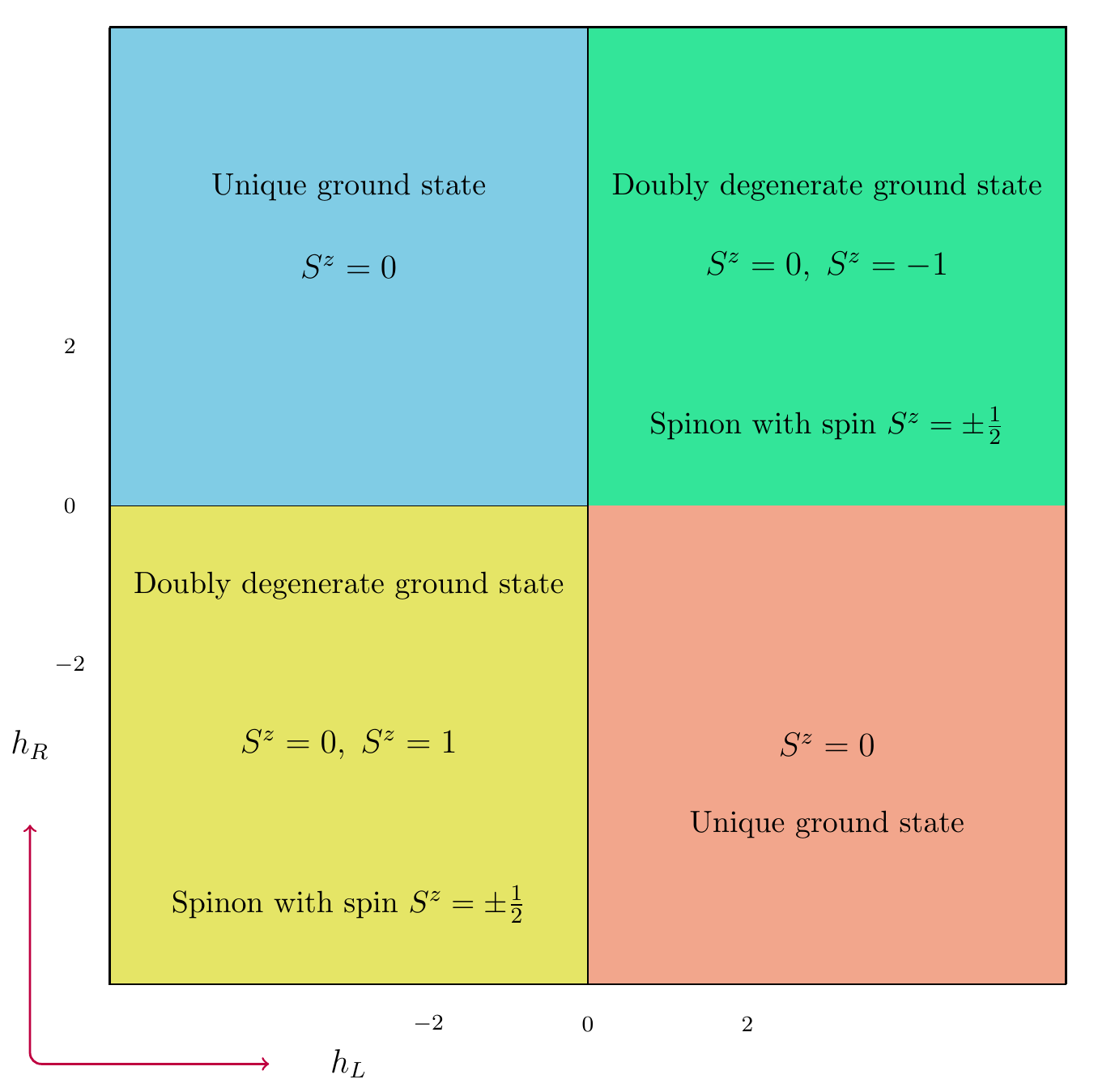}
\caption{The figure shows various ground states occurring for different values of the boundary magnetic fields. The ground state is unique in the second and fourth quadrants. In the first and third quadrants, the ground state contains a spinon with infinite rapidity and whose spin is oriented either in the positive or negative $z$ direction, resulting in a two fold degenerate ground state.}   
\label{fig:evengs}
\end{figure}
\end{center}

We first discuss the ground state properties of the model. We present in the Fig. (\ref{fig:evengs}) the ground state phase diagram in the plane $(h_L, h_R)$ and for an even number of sites. A similar analysis can be made for an odd number of sites as given in section (\ref{sec:gsBethe}). There are four  different possible ground states when  labelled by  the conserved total ``$z$'' component of the spin operator, 
\be
S^z=\frac{1}{2}\sum_{j=1}^N \sigma_j^z. 
\label{Sz}
\ee
When
$h_L h_R < 0$ the ground state is unique and has total  spin $S^z=0$ whereas, contrarily to what was found in Ref.\cite{kapustinxxz}, in the quadrants $h_L h_R > 0$ we find that the ground state is {\it doubly degenerated}; each one having spin $S^z=0$ and $S^z=+ 1$ or $S^z=- 1$ depending on whether $h_{L(R)}$ is negative or positive. In the later cases, the degeneracy is found to be the consequence of the existence of two static spinons (with infinite rapidity) with spins $\pm \frac{1}{2}$ in the ground state. To get a better understanding of the ground state structure we have performed extensive density matrix renormalization group (DMRG) calculations and calculated the magnetization profile $\langle S^z_j \rangle$ in the ground state. Overall we find that the edge magnetic fields induce a spin polarization close to the two edges which extends into the bulk in a power law fashion. Furthermore,  we find that the corresponding spin accumulations at the edges are {\it fractional} and take the values $\pm \frac{1}{4}$ (opposite to  the orientation of the edge field)  at large fields $|h|> 2$. The  situation at hand   is similar to what happens in  gapless symmetry protected topological (SPT) superconductors \cite{Keselman2015,Keselman2018,PAA1,PAA2} where the edge states Hilbert space is exhausted by  spin-$\frac{1}{4}$ operators. However in the present case, due to the existence of massless spinon bulk excitations, such an operator do  not represent a genuine fractional  sharp quantum observable  since, as  we show in the section (\ref{sec:numerical}),   its variance is not zero. 

\begin{center}
\begin{figure}[!h]
\includegraphics[width=1\columnwidth]{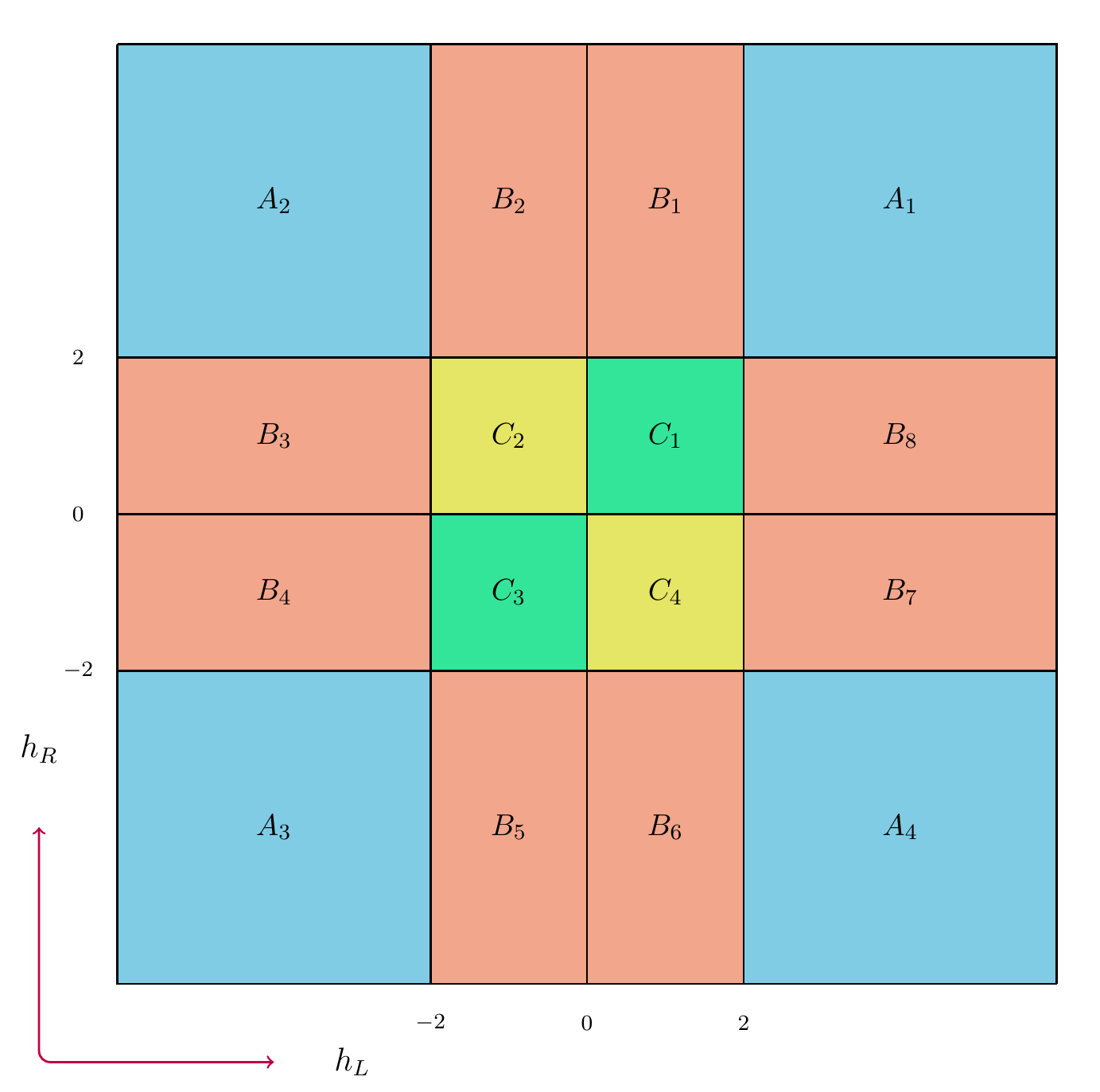}
\caption{The figure shows various phases occurring for different values of the boundary magnetic fields. The $\it{A}$ sub-phases exhibit two boundary bound-states, one at each edge. In $\it{B}$ sub-phases there exists one boundary bound state at either the left or the right edges. In $\it{C}$ sub-phases boundary bound-states do not exist. 
\label{fig:PD}}
\end{figure}
\end{center}

The second topic we shall discuss in this work concerns the structure of the excited states and its relation to the existence of bound states localized at the edges.
One of the hallmarks of the boundary physics induced by the edge magnetic fields is the existence of boundary bound states localized close to the edges where the fields are applied. As previously found in Refs.\cite{skorik,kapustinxxz},
when the magnitude of fields are large enough, i.e: when $|h_{L(R)}| \ge 2$, the system hosts  bound states with energy
\bea \label{bounden}
 m_{L,R}= 2\pi/\sin(\frac{\pi}{h_{L,R}}), 
 \eea
which carry  a spin $\frac{1}{2}$, whose spin orientation is {\it along} the boundary fields at each edge. Contrarily to the zero energy edge states in SPT massless superconductors, the bound states in the XXX model  are high-energy states whose energies are always above the one spinon branch of massless bulk excitations, and as we shall see, their existence has important consequences on the structure of the Hilbert space. We present in  Fig. \ref{fig:PD} the bound state phase diagram of the model (\ref{eq:H}). In each quadrant the different phases are sorted out as a function of the number of bound states: the $A$ sub-phases support two bound states (one at each edge), the $B$ sub-phases support one bound state at either the right or the left edge,  whereas in the $C$ sub-phases there are no bound states. When compared to the ground state phase diagram we see that each quadrant is split into three different sub-phases named $A$, $B$ or $C$. Although in all these sub-phases the ground state has the same total spin $S^z$, they  differ by the structure of the high energy states. We show that  each bound state generates a whole tower of excited states that can be built upon it. Hence, the Hilbert space is comprised of a certain number of towers which depends on the number of bound states exhibited by the system. In the case of the gapped regime of the spin $\frac{1}{2}$ XXZ spin chain with open boundaries a similar structure of the Hilbert space is found, where it was demonstrated in \cite{Fendley} that the Hilbert space is comprised of two towers of degenerate eigenstates which leads to the emergence of a strong zero energy Majorana operator (which commute with the Hamiltonian) which map these pairs of states.

In each of the $A, B, C$ sub-phases, the direct sum of these towers span the complete Hilbert space. When crossing the boundaries between any two of these sub-phases, since the number of bound states exhibited by the system and hence the number of towers of the excited states changes, an Eigenstate Phase Transition occurs which involves a full reorganization of the Hilbert space. A similar phenomenon is observed in gapless SPT superconductors \cite{PAA2}.

In summary, we find that similar to the systems exhibiting SPT \cite{PAA2} the XXX spin chain exhibits several phases as a function of edge magnetic fields. We find that the total spin of the ground state is not enough to completely characterize these phases which differ also by the structure of the Hilbert space. The later is  linked  to the number of bound states at the edges which generate towers of excited states that together span the Hilbert space. As a consequence, on top  of the    phase transition corresponding to a  change in the ground state, there exists eigenstate phase transitions involving the change in the number of towers of the Hilbert space.

The paper is organized as follows. We present our results obtained from the Bethe ansatz   for  the ground state  and the excited states in the sections \ref{sec:gsBethe} and \ref{sec:excitedBethe} respectively. Section \ref{sec:numerical} is dedicated to the DMRG analysis of the ground state properties. We finally discuss our results in the section \ref{sec:discussion}.

\section{ Ground state phase diagram}
\label{sec:gsBethe}

As said above, the Hamiltonian (\ref{eq:H}) is integrable by the method of the Algebraic Bethe ansatz \cite{GrisaruMezincescNepomechie,ODBA} for arbitrary values of the boundary fields. Its ground-state as well as excitations are obtained from the Bethe equations 
 \bea \label{Bethe}
\left(\frac{\lambda_j-i/2}{\lambda_j+i/2}\right)^{2N}\left(\frac{\lambda_j+i(\frac{1}{2}-p_{L})}{\lambda_j-i(\frac{1}{2}-p_{L})}\right)\left(\frac{\lambda_j+i(\frac{1}{2}-p_{R})}{\lambda_j-i(\frac{1}{2}-p_{R})}\right)\nonumber\\=\prod_{j\neq k=1}^{M}\left(\frac{\lambda_j-\lambda_k-i}{\lambda_j-\lambda_k+i}\right)\left(\frac{\lambda_j+\lambda_k-i}{\lambda_j+\lambda_k+i}\right),\eea
where we have introduced  $p_{L/R}=1/h_{L/R}$ as the boundary parameters. The eigenstates of the Hamiltonian are labelled by $M \in \mathbb{N}$  Bethe roots $\lambda_{j=1,...,M}$ which are solutions of Eq.(\ref{Bethe})
and have energy
\bea 
 E=-\sum_{j=1}^{M}\frac{2}{\lambda_j^2+\frac{1}{4}}+N-1+ h_L + h_R.
 \eea
 The corresponding total spin $S^z$ of a state is related to the integer $M$ through the relation (\ref{nrootsH})  \bea \label{nrootsH}S^z= \pm \left(\frac{N}{2}-M\right).\eea 
Where $\pm$ corresponds to reference state with all spin up and down respectively. We have obtained from (\ref{Bethe}) the ground state phase diagram as a function of the boundary magnetic fields $h_{L(R)}$ and for both an even and an odd number of sites. Before going into more details let us first review briefly the situation at zero fields. In this case the ground state depends on the parity of $N$ as follows: For even $N$ it is non degenerated and has total spin $S^z=0$ whereas for odd $N$ it is twofold degenerated, each ground state having total spin $S^z=\pm \frac{1}{2}$. The latter degeneracy is due to the presence of spin-$\pm \frac{1}{2}$ spinons,  which have  energy \bea\label{energyhole} E_{\theta}=\frac{2\pi}{\cosh(\pi\theta)}, \eea
and is zero in the limit of  infinite rapidity $\theta \rightarrow \infty$. Let us see now how this scheme is modified by the presence of non zero boundary fields. The situation further depends on the parity of $N$.


\subsection{Ground state for odd number of sites}
\begin{center}
\begin{figure}[!h]
\includegraphics[width=1\columnwidth]{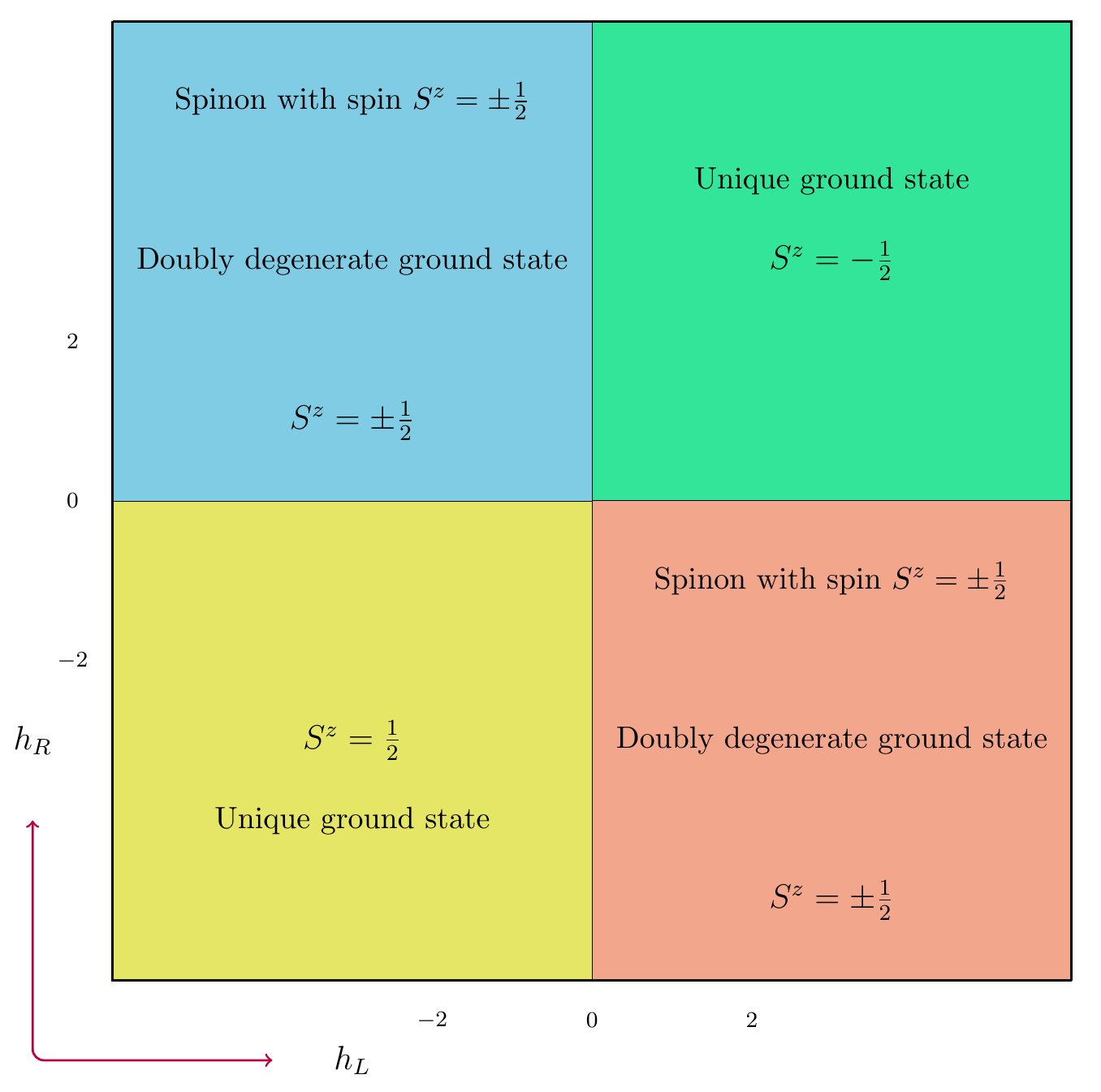}
\caption{Ground state phase diagram as a function of the boundary fields for an odd number of sites. 
The ground state is unique in the first and third quadrants. In the second and fourth quadrants, the ground state contains a spinon with infinite rapidity and whose spin is oriented either in the positive or negative $z$ direction, resulting in a two fold degenerate ground state.}   
\label{fig:oddgs}
\end{figure}
\end{center}

Since the total number of sites is odd the spins of the ground states have to be half integers.
The phase diagram can be broadly divided into four quadrants based on the direction of the boundary magnetic fields as shown in the Fig. (\ref{fig:oddgs}). In the upper right quadrant, when both the magnetic fields point towards the positive $z$ direction, and independently of their magnitudes $|h_{L(R)}|$,  the ground state is unique and has a total spin $S^z= -\frac{1}{2}$. In the ground state, that we shall label 
\be
|-\frac{1}{2} \rangle,
\ee
the total magnetization is due to a static spin configuration which account for the total spin $-\frac{1}{2}$. In the lower right quadrant, in which $h_L >0$ and $h_R <0$, the ground state is doubly degenerated and carry total spins $S^z= \pm \frac{1}{2}$
\be
|-\frac{1}{2} \rangle \;{\rm and}\;  |+\frac{1}{2} \rangle
\ee
In contrast with the previous case, the spins of the ground states here is due to the presence of spin-$\pm \frac{1}{2}$ spinons  with  infinite rapidity $\theta \rightarrow \infty$ (\ref{energyhole}).  The situation in the two other quadrants, i.e. the lower left and upper left ones, can be obtained by using the isometry (\ref{z2}) and reversing the total spin quantum number $S^z \rightarrow -S^z$. The ground states  are then found to be 
\be
|+\frac{1}{2} \rangle \; {\rm and}\; |- \frac{1}{2} \rangle,
\ee
in the upper left quadrant, i.e. when  $h_L < 0$ and $h_R > 0$ and
\be
|+\frac{1}{2} \rangle,
\ee
in the lower left quadrant when $h_L < 0$ and $h_R < 0$.

\subsection{Ground state for even number of sites}
In this case the spins of the ground states are always integers. As shown in the Fig.(\ref{fig:evengs}) in the upper right quadrant, i.e. when both $h_{L(R)} >0$, the ground state is doubly degenerated and have total spins $S^z=0,-1$ represented by
\be
|0\rangle \, {\rm and} \;  |-1\rangle.
\ee
The double degeneracy of the ground state is due to the presence of spin-$\pm \frac{1}{2}$ spinons
with infinite rapidity. These spinons have to be  added {\it on top} of a background static spin configuration contributing to a total spin $-\frac{1}{2}$ in such a way that the total spins  of the ground states are integers $S^z=0,-1$. Using the isometry (\ref{z2}) we deduce immediately that in the lower left quadrant the ground states are given by
\be
|0\rangle \, {\rm and} \;  |+1\rangle,
\ee
and also contain spin-$\pm \frac{1}{2}$ spinons
with infinite rapidity on top of a background spin $\frac{1}{2}$ configuration.
This is to be true independently of the magnitudes of the fields $|h_{L(R)}]$. 
When the boundary fields point towards opposite direction, like in the two upper left and lower right quadrants the ground state is unique with  total spin $S^z=0$,
\be
|0\rangle,
\ee
and do not contains spinons.

As we have seen the ground state phase diagram exhibits four distinct phases  depending solely on the orientations of the boundary fields. In each of the four quadrants 
defined by the sign of $h_L$ and $h_R$ the ground state degeneracy depends on the parity of $N$. It is two-fold degenerate when $h_L h_R <0$ for $N$ odd and when $h_L h_R >0$ for $N$ even. In all other cases the ground state is non degenerate contrary to what was found in \cite{kapustinxxz}.  Overall, our understanding of the spin quantum numbers in the different phases relies on a static background spin distribution on top of which spins $\pm \frac{1}{2}$ spinons may or may not be added. Independently of the parity of the number of sites $N$, the background spin distribution contributes  to a total spin $S^z_B= -\sgn(h_L) \;  \frac{1}{2}$  when $h_{L}h_{R} > 0$ whereas  $S^z_B= 0$ in the opposite case when $h_{L}h_{R} < 0$. Such a background spin structure is   due to the presence of the boundary fields  $h_{L},h_{R}$ which are expected
to induce a spin accumulation close to the edges. We shall return to this point  
in the section (\ref{sec:numerical}) when we shall study the ground state properties in more detail. For the time being we shall argue that the phase structure  induced by the presence of the boundary fields is much richer than the one we have just presented. When considering the whole structure of the Hilbert space, which calls for a  detailed description of the excited states,  we shall show that  each of the four quadrants, $(h_L \gtrless0, h_R\gtrless 0)$, splits into four distinct 
sub-phases where the excited states organizes into different towers. This is the consequence of the well-known fact that the edge fields, when large enough, induce boundary bound states that are exponentially localized close to the edges.

\section{Excited states}
\label{sec:excitedBethe}

Before going into the description of the full solution of (\ref{hamiltonian}), let us review quickly how, in the case of a periodic chain without fields, the structure of the excitation  spectrum depends on the parity of the number of sites $N$. 
When $N$ is {\it even} the ground-state is a singlet
and the  excitations are obtained by   adding an {\it even} number of propagating holes or spinons.  The spinons carry spins $\pm \frac{1}{2}$ and have energy \eqref{energyhole}. Since  physical excitations  correspond to flipping a certain number of spins in the chain they  carry integer spins $S^z \in  \mathbb{Z}$ and therefore the spinons
 always come in pairs in an even chain. 
For the spin chain with an {\it odd} number of sites, the ground state is two-fold degenerate with total spins $S^z=\pm\frac{1}{2}$. Each of the ground-states   contains a spinon, with zero energy in the thermodynamical limit,  and  rapidity $\theta\rightarrow \infty$. In contrast with the even chain case, single spinon excitations with a finite rapidity, i.e: $\theta \neq \infty$, and  energy (\ref{energyhole}) are allowed when $N$ is odd. All other  excitations are then obtained by adding  an even number of spinons to the above states. Hence  the total number of spinons in the odd spin chain  is always an odd integer.

This scheme is to be modified in an open chain with boundary fields for which  the ground-state and, more importantly, the very  structure of the Hilbert space of excitations   strongly depend on the boundary  fields $h_{L/R}$. As we shall see, when the boundary fields are strong enough (i.e. when $|h_{L/R}| \ge 2$) their main  effect is to stabilize bound-states which are {\it localized} at either the left or the right edge. These bound-states have a finite, i.e. non zero, energy above the ground-state, 
\be 
\label{boundensum}
 m_{L,R}= \frac{2\pi}{\sin(\frac{\pi}{h_{L,R}})}, 
 \ee
and carry a spin $\frac{1}{2}$ which points {\it towards}
the direction of the boundary field at each edge. The bound states at the left and the right edges are exponentially localized as $\sim e^{-\kappa_L x}$ and $\sim e^{-\kappa_R (N-x)}$ respectively (see Appendix \ref{appendix2}), where \be \kappa_j=\log \left(\frac{h_j+1}{h_j}\right), \;\;\; j=L,R. \ee

When the bound states exist, they generate independent towers of excited states on top of the ground-state one. All these towers of states eventually span the whole Hilbert space. 

We distinguish between three regions, or sub-phases, $A, B$ and $C$ depending on the number of localized bound-states in the spectrum. In the region $A$ both boundary field strengths exceed a critical value $|h_{L/R}| \ge 2$ and there exists  two boundary bound-states localized at both ends of the chain. Depending on the relative orientations of the fields $h_{L/R}$ with respect to the "z" axis we further distinguish between  four sub-phases $A_{j=(1,2,3,4)}$.  In the region $B$ only one boundary field strength exceeds the critical value and there exists a single  bound-state which is localized at either the left or the right edge. Taking into account the orientations of the fields we end up with eight sub-phases $B_{j}$ $j=(1...8)$. Finally  in region $C$, both $|h_{L/R}| < 2$ and there are no localized bound-states; the four sub-phases $C_{j=(1,2,3,4)}$ account for  all possible boundary fields orientations. 
The phase diagram is depicted in Fig.(\ref{fig:PD}).

In the following  we shall present 
our results  for the ground-states as well as the Hilbert space structures in  each phase. Since, as with the PBC case discussed above, the spectral properties are very sensitive to the evenness of the number of sites $N$, we shall discuss separately  both even and odd chains.



\subsection{$\it{A}$ sub-phases}
\label{sec:Aphase}
We start with the $A$ sub-phases where two boundary bound-states are stabilized.
The four $A_{j=(1,2,3,4)}$ sub-phases  correspond to the domains of boundary fields 
 $(h_{L}\ge 2,h_{R}\ge 2), (h_{L} \le -2,h_{R}\ge 2) ,(h_{L} \le -2,h_{R}\le -2)$ and $(h_{L} \ge 2,h_{R}\le -2)$ respectively. In the following we shall distinguish  between odd end even chains and discuss separately 
 the sub-phases $A_{j=(1,3)}$ and $A_{j=(2,4)}$.
 
\subsubsection{Odd number of sites}
\label{sec:A1oddnumber}
\paragraph{The $A_1$ and $A_3$ sub-phases.}
In  these cases  both  boundary magnetic fields point
towards the same direction: along  the positive  $z$ axis
for  the  $A_1$ sub-phase and negative $z$ axis for 
the  $A_3$ sub-phase. Both cases are related by the isometry (\ref{z2}). Qualitatively speaking, in the sub-phases $A_{1,3}$ and for $N$ odd, the boundary magnetic fields are not frustrating in the sense that 
in the Ising  limit of (\ref{hamiltonian})  the ground-state would exhibit perfect antiferromagnetic order.

In the $A_1$ sub-phase we find that the ground-state is unique and has a total spin $S^z=-\frac{1}{2}$.
We accordingly label the ground-state in this phase   by
\be\label{gsA1}
|-\frac{1}{2}\rangle,
\ee
and denote by $E_0$ its energy. The expression of $E_0$  as a function of $h_{L,R}$ is given  in the Appendix (see Eq.(\ref{energygroundAdown})). 
We notice that due to the presence of the boundary fields the spin $-\frac{1}{2}$ of the ground-state is not carried by a spinon in contrast with the periodic chain with $N$ odd. It is rather the consequence of a static spin density
distribution. We shall discuss this topic in more detail
in the next section.
Similarly  to the case of periodic boundary conditions, one can build up excitations in the bulk on top of this ground state by adding an arbitrary {\it even} number of spinons, bulk strings and quartets \cite{DestriLowenstein}. These bulk excitations 
 built on top of the state $|-\frac{1}{2}\rangle$  form a tower of excited states 
 that we shall denote the ground-state tower.

As said above in the $A$ sub-phases there exists two boundary bound-state solutions exponentially
localized at either the  left or the right edge.
In the language of the Bethe ansatz they correspond to purely imaginary solutions of (\ref{Bethe}) (see Appendix \ref{appendix1}). These bound-states carry  a spin $\frac{1}{2}$, whose spin orientation is along the boundary fields
 at each edge, and have an energy \eqref{bounden}
Since the bound-states carry  a spin half, in order  to add a bound-state  to the ground-state 
one also needs to add a spinon.  This spinon may have  spin $+\frac{1}{2}$ or $-\frac{1}{2}$ and an arbitrary 
 rapidity $\theta$. The energy cost in the process is
 \be
 \label{bsspinon}
 E_0+m_{L,R}+E_{\theta},
 \ee
 and is minimal when $\theta \rightarrow \infty$.
 The corresponding states  
 \be
 \label{LRtowers}
|\pm \frac{1}{2}\rangle_{L}\; {\rm and}\; |\pm \frac{1}{2}\rangle_{R}\; ,
 \ee
 have total spins $S^z=\pm \frac{1}{2}$ and energies $E_0+m_{L}$ and  $E_0+m_{R}$. 
 The lowest excited states above (\ref{LRtowers}) consist of spinon branches with energies given by (\ref{bsspinon}) and $\theta \neq \infty$. On top of these, the states (\ref{LRtowers})
generate, each, a tower of excited states obtained by  adding an arbitrary {\it even} number of spinons, bulk strings and quartets. In both the left and right towers, built upon (\ref{LRtowers}), a localized bound-state at the left and the right edge is present and the number of spinon excitations is always odd. 

\begin{center}
\begin{figure}[!h]
\includegraphics[width=0.9\columnwidth]{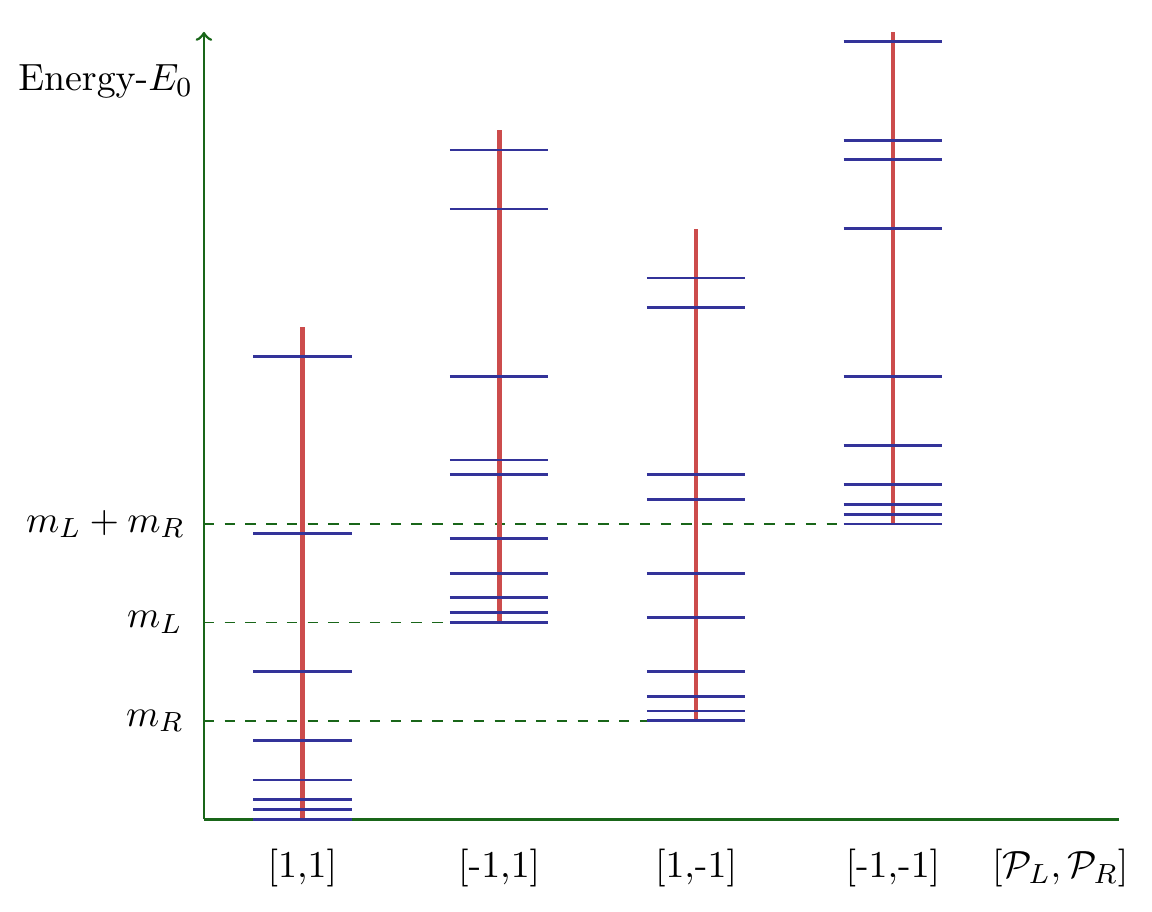}
\caption{The figure illustrates the structure of the Hilbert space in the $A$ sub-phases. In each $A_j$ ($j=1,2,3,4$) sub-phase, there exist four towers of excited states, each labelled by the bound state parities [$\mathcal{P}_L$, $\mathcal{P}_R$]. The lowest energy states in each tower have energies close to the dashed lines corresponding to the energies $E_0$, $E_0+m_L$, $E_0+m_R$ and $E_0+m_L+m_R$. 
\label{fig:hilberttowers}}
\end{figure}
\end{center}

On top of the above three towers there exists a fourth one which correspond to states which host two bound-states. The state with the lowest energy in this tower is obtained by adding a localized bound-state at the left and the right edges to the ground-state (\ref{gsA1}). Since in the process the total spin of the state is shifted by $1$, no spinon is required. The resulting state 
\be\label{L&Rtower}
| +\frac{1}{2}\rangle_{LR}\; ,
\ee
which has a total spin $S^z=\frac{1}{2}$ and an energy 
$E_0+m_{L}+m_{R}$, generates a tower of excited states
that comprises an arbitrary even number of spinons, bulk strings and quartets. The number of spinon states in the whole tower is always even. We thus see that, in the $A_1$ sub-phase, the whole Hilbert space can be split  into four towers generated by the states (\ref{gsA1}, \ref{LRtowers}) and (\ref{L&Rtower}) as illustrated in figure \ref{fig:hilberttowers}. On top of the ground-state tower which governs the low-energy physics,
the remaining three towers contain at least one bound-state at the edges and are high-energy states.
In particular, we notice that in the $A_1$ sub-phase there exists excitations which contain a single spinon, and although the system is massless, their minimum energy is greater than the boundary gap $m_L$ or $m_R$. These four towers  can be labelled by the bound state parities 
\be
\mathcal{P}_{L,R}=(-1)^{\mathcal{N}_{L,R}},
\label{bsparity}
\ee
where $\mathcal{N}_{L,R}$ correspond to number of bound states at the left and right edges respectively.

The situation in the $A_3$ sub-phase can be described in the very same way as above. Using the isometry
(\ref{z2}), we can obtain all the states in the sub-phase $A_3$ starting from the states in the sub-phase $A_1$ by reversing the sign of the total spin $S^z$ of the states.
Hence, we obtain four towers of states in the sub-phase $A_3$ generated by the states $| +\frac{1}{2}\rangle$, $| \pm \frac{1}{2}\rangle_{L,R}$ and $| -\frac{1}{2}\rangle_{LR}$
at energies $E_0, E_0+m_{L,R}$ and $E_0+m_{L} +m_{R}$.

\paragraph{The  $A_2$ and $A_4$  sub-phases.}
\label{sec:A2oddnumber}
In these cases the boundary fields are frustrating for $N$ odd  in the sense discussed above. 
As we shall see in these sub-phases the Hilbert space is also split into four towers of states corresponding to the presence of  boundary bound-states.  
However, since  the boundary magnetic fields at the two edges point toward opposite directions, the nature of these towers differ from the ones described above. 
Consider for instance the $A_2$ sub-phase in which 
the left boundary field points towards the negative $z$ axis  while   the one at the right boundary points in the opposite direction. In this case we find that the ground-state is two-fold degenerated, each one  containing a spinon (but no bound-state)  with spin 
$\pm\frac{1}{2}$ and rapidity $\theta\rightarrow\infty$.
These two states, i.e:
\be\label{gsA2}
|\pm\frac{1}{2}\rangle,
\ee
 have total spin $S^z=\pm\frac{1}{2}$ corresponding to the spin  of the spinon, and generate a tower of excited state. It is obtained by adding an arbitrary even number of spinons, bulk strings and quartets 
 on top of the two  spin $\pm\frac{1}{2}$ massless  spinons  branches with spectrum (\ref{energyhole}) and rapidity $\theta \neq \infty$. In contrast with the $A_{1,3}$ sub-phases the ground-state tower contains an odd number of spinons. 

Just as in the sub-phase $A_1$, there exists two boundary bound-state solutions one at each edge.  The bound state's spin is always oriented along  the boundary magnetic field. Hence, in the sub-phase $A_2$ the bound-state localized at the left edge has spin $-\frac{1}{2}$ whereas the bound-state localized at the right edge has spin $+\frac{1}{2}$. We find that in order to add the bound-state at the left edge with spin $-\frac{1}{2}$  one has to remove the
 $-\frac{1}{2}$ spinon at $\theta=\infty$ in the $|-\frac{1}{2}\rangle$ ground-state (\ref{gsA2}).
 The resulting state  has  total spin $S^z=-\frac{1}{2}$ and energy $E_0+m_L$. Similarly adding a spin $+\frac{1}{2}$ bound-state at the right edge requires to remove the  spin $\frac{1}{2}$ spinon from the ground-state $|+\frac{1}{2}\rangle$  (\ref{gsA2}). 
 The resulting state  has  total spin $S^z=+\frac{1}{2}$ and energy $E_0+m_R$. The two states with a bound-state at either the left or right edge
 \be\label{LRA2towers}
 |-\frac{1}{2}\rangle_L \; {\rm and}\; |+\frac{1}{2}\rangle_R,
 \ee
 generate, each, a tower of excited states upon adding 
 an arbitrary even number of spinons, bulk strings and quartets. In these two towers the number of spinons in every state
 is always even.
 
 Finally, the fourth tower is obtained by adding a bound-state at each edge to  the two ground-states (\ref{gsA2}). The total spin of the resulting state 
 does not change since the two, left and right, bound-states have opposite spins. We obtain the states
 \be\label{LandRA2towers}
 |\pm \frac{1}{2}\rangle_{LR} ,
 \ee
which have an energy $E_0+m_L+m_R$ and generate a tower
of excited states. It is obtained by adding even number of spinons, bulk strings and quartets. In this tower the number of spinons is always odd.

Using the symmetry \ref{z2}, we can obtain all the states in the sub-phase $A_4$ from the states in the sub-phase $A_2$  by reversing their spins.
The Hilbert space in the 
sub-phase $A_4$ can be similarly sorted out in terms of four towers of states built upon the states 
$|\pm\frac{1}{2}\rangle$, $|+\frac{1}{2}\rangle_L,  |-\frac{1}{2}\rangle_R$ and $|\pm \frac{1}{2}\rangle_{LR}$
with energies $E_0, E_0+m_{L,R}$ and $E_0+m_{L} +m_{R}$.

 \begin{table}[h!]
\centering
\caption{Energies and bound state parities of the ground state and the lowest energy states corresponding to each tower in all the $A$ sub-phases for odd number of sites is shown below. The subscripts $L,R$ denote the location of the bound states at the left or the right boundary.}

\begin{tabular}{|c|c|ccc|}
\hline
\hline
\;\;Phase\;\;&\;\;State\;\;&\;\;Energy-$E_0$\;\;&\;\;$\mathcal{P}_L$\;\;&\;\;$\mathcal{P}_R$\;\;\\
\hline
\;&$|-\frac{1}{2}\rangle$& $0$ \;(g.s)& 1 & 1\\
\;&$|\pm\frac{1}{2}\rangle_{R}$& $m_R$ & 1 & -1\\
$A_1$&$|\pm\frac{1}{2}\rangle_{L}$& $m_L$ & -1 & 1\\
\;&$|\frac{1}{2}\rangle_{L,R}$&  $m_L+m_R$& 1 & 1\\

\hline

\;&$|\frac{1}{2}\rangle$& $0$ \;(g.s)& 1 & 1\\
\;&$|\pm\frac{1}{2}\rangle_{R}$& $m_R$ & 1 & -1\\
$A_3$&$|\pm\frac{1}{2}\rangle_{L}$& $m_L$ & -1 & 1\\
\;&$|-\frac{1}{2}\rangle_{L,R}$&  $m_L+m_R$& 1 & 1\\

\hline

\;&$|-\frac{1}{2}\rangle_{L}$& $m_L$& -1 & 1\\
\;&$|\pm\frac{1}{2}\rangle$& $0$ \;(g.s) & 1 & 1\\
$A_2$&$|\pm\frac{1}{2}\rangle_{L,R}$& $m_L+m_R$ & -1 & -1\\
\;&$|\frac{1}{2}\rangle_{R}$&  $m_R$& 1 & -1\\

\hline

\;&$|-\frac{1}{2}\rangle_{R}$& $m_R$& 1 & -1\\
\;&$|\pm\frac{1}{2}\rangle$& $0$ \;(g.s) & 1 & 1\\
$A_4$&$|\pm\frac{1}{2}\rangle_{,L,R}$& $m_L+m_R$ & -1 & -1\\
\;&$|\frac{1}{2}\rangle_{L}$&  $m_L$& -1 & 1\\
\hline
\hline
\end{tabular}
\label{tableAodd}
\end{table}

\subsubsection{Even number of sites}

When the number of sites is even the frustrating effect of the magnetic fields is reversed as compared to the $N$ odd
case. The boundary fields are frustrating in sub-phases $A_{1,3}$ while non-frustrating in the sub-phases $A_{2,4}$.

\paragraph{The   $A_1$ and $A_3$ sub-phases.}
\label{sec:A1evnnumber}
In the sub-phase $A_1$ we find that the ground-state
is two-fold degenerated. It does not contain bound-states
but does contain a spinon with  rapidity 
$\theta \rightarrow \infty$ and spins $\pm \frac{1}{2}$.
Despite this, since $N$ is even, the total spins of the two degenerate ground-states have to be  integers. Indeed,
as it comes out from our exact solution the two ground-states have total spins $S^z=0$ and $S^z=-1$. Our interpretation of this fact is that 
the two ground-states with spin $S^z=0$ and $S^z=-1$ contain a spin $+\frac{1}{2}$ and  a spin $- \frac{1}{2}$ spinon respectively on top of a static background spin $- \frac{1}{2}$ distribution corresponding to the ground state in $A_1$ sub-phase  when $N$ is odd. In the following we denote these two ground-states by
\be\label{evenA1gs}
|0\rangle\;  {\rm and} \; |-1\rangle. 
\ee
The ground-state tower of excitated states comprises
spin $\pm \frac{1}{2}$ massless spinon states with energy
$E_0+m_{R}+E_{\theta}$ and finite rapidity $\theta \neq \infty$. The rest of the tower 
is then obtained by adding an arbitrary even number of spinons, bulk strings and quartets. In this tower the number of spinon states is always odd.

Starting from one of the two ground-states (\ref{evenA1gs}), one may add a  bound-state 
at either the left or the right edge. To this end one needs to remove the spin $\pm \frac{1}{2}$  spinon. The resulting total spin is then the sum of the bound-state spin  $+\frac{1}{2}$ with that of the static background spin $- \frac{1}{2}$ distribution mentioned above. 
As a result, we end up with two states of  total spin $S^z=0$. The corresponding states with the bound-state at the left or the right  edge are denoted
\be\label{evenA1bstowers}
|0\rangle_L \; {\rm and}\; |0\rangle_R,
\ee
and have energies $E_0+m_{L}$ and $E_0+ m_{R}$. Each of these two states generates a tower of excited states.
In these towers the number of spinon states is always even.

The fourth tower is obtained from the ground-states (\ref{evenA1gs}) by adding a bound-state at each edge. 
Since the change of total spin is $1$ there is no need to add or remove a spinon. In the process we obtain two degenerate states, with total spins $S^z=1$ and $S^z=0$
and energy  $E_0+m_L+m_R$, 
\be\label{evenA1LRtowers}
|1\rangle_{LR} \; {\rm and}\; |0\rangle_{LR},
\ee
that host spin $\pm \frac{1}{2}$ spinons with infinite rapidity as in the ground-states. The fourth tower of excited states comprises, as in the ground-state tower, spin $\pm \frac{1}{2}$  spinon states. These states  have energy $E_0+m_{L} +m_{R}+E_{\theta}$ and are gapped high energy states. The remaining states of this towers 
are then built up  by adding an even number of spinons, 
bulk strings and quartets. The number of spinon states is always odd.

Similar to the odd number of sites case, using the symmetry (\ref{z2}), we can obtain all the states in the phase $A_3$ starting from the states in the phase $A_1$ described above.

\paragraph{The   $A_2$ and $A_4$ sub-phases.}
\label{sec:A2evnnumber}
 
In the sub-phase $A_2$  we find that the ground-state
is non-degenerated
\be\label{A2evengs}
|0\rangle,
\ee
 and has total spin $S^z=0$ with  energy $E_0$. Starting from this ground state we can add a bound-state at the left edge with spin $-\frac{1}{2}$. As already emphasized one also needs to add a spinon, with infinite rapidity and zero energy, for the total spin shift to be an integer.  Depending on the spinon spin, which can be either $\pm \frac{1}{2}$, one ends up with two states
 \be\label{A2evenLtower}
 |-1\rangle_{L},|0\rangle_{L},
 \ee
 which have total spins $S^z=-1$ and $S^z=0$ and energy 
 $E_0+m_L$. One may repeat the same line of arguments with the right edge paying attention that the bound-state's spin in this case is $+\frac{1}{2}$. The resulting two states
 \be\label{A2evenRtower}
 |+1\rangle_{R},|0\rangle_{R},
 \ee
 hosting a bound-state at the right edge have total spins
 $S^z=1$ and $S^z=0$ and energy $E_0+m_R$. Each left and right states (\ref{A2evenLtower}) and (\ref{A2evenRtower}) generate two towers of excited states that comprise spin $\pm \frac{1}{2}$ spinons 
 with energies $E_0+m_{L,R}+ E(\theta)$. The rest of the towers are obtained by adding even number of spinons, bulk strings and quartets.
 
The forth tower is obtained from the ground-state (\ref{A2evengs}) by adding a bound-state with spin
$- \frac{1}{2}$ at the left edge and spin
$+ \frac{1}{2}$ at the right edge. No spinons are needed
in the process and one ends up with a single state
`\be\label{A2evenLRtower}
 |0\rangle_{LR},
 \ee
with total spin $S^z=0$ and energy $E_0+m_R+m_L$. The latter state generates also a tower of states with even number of spinons, bulk strings and quartets.

Using the symmetry (\ref{z2}), similar to the odd number of sites case, we can obtain all the states in the sub-phase $A_4$ starting from the states in the sub-phase $A_2$ described above. The ground state and the lowest energy state corresponding to each tower in all the $A$ sub-phases for odd and even number of sites chain are summarized in the tables \ref{tableAodd} and \ref{tableAeven} respectively. 

\begin{table}[h!]
\centering
\caption{Energies and bound state parities of the ground state and the lowest energy states corresponding to each tower in all the $A$ sub-phases for even number of sites is shown below.}

\begin{tabular}{|c|c|ccc|}
\hline
\hline
\;\;Phase\;\;&\;\;State\;\;&\;\;Energy-$E_0$\;\;&\;\;$\mathcal{P}_L$\;\;&\;\;$\mathcal{P}_R$\;\;\\
\hline
\;&$|-1\rangle, \; |0\rangle$& 0 \;(g.s)& 1 & 1\\
\;&$|0\rangle_{R}$& $m_R$ & 1 & -1\\
$A_1$&$|0\rangle_{L}$& $m_L$ & -1 & 1\\
\;&$|1\rangle_{L,R}, \; |0\rangle_{L,R} $&  $m_L+m_R$& 1 & 1\\

\hline

\;&$|1\rangle, \; |0\rangle$& 0 \;(g.s)& 1 & 1\\
\;&$|0\rangle_{R}$& $m_R$ & 1 & -1\\
$A_3$&$|0\rangle_{L}$& $m_L$ & -1 & 1\\
\;&$|-1\rangle_{L,R},\; |0\rangle_{L,R}$&  $m_L+m_R$& 1 & 1\\

\hline

\;&$|0\rangle$& 0 \;(g.s)& -1 & -1\\
\;&$|-1\rangle_{L}, \; |0\rangle_{L}$& $m_L$ \; & 1 & -1\\
$A_2$&$|1\rangle_{R}, \; |0\rangle_{R}$& $m_R$ \; & -1 & 1\\
\;&$|0\rangle_{L,R}$&  $m_L+m_R$& 1 & 1\\

\hline

\;&$|0\rangle$& 0 \;(g.s)& -1 & -1\\
\;&$|1\rangle_{L}, \; |0\rangle_{L}$& $m_L$ \; & 1 & -1\\
$A_4$&$|-1\rangle_{R}, \; |0\rangle_{R}$& $m_R$ \; & -1 & 1\\
\;&$|0\rangle_{L,R}$&  $m_L+m_R$& 1 & 1\\
\hline
\hline
\end{tabular}
\label{tableAeven}
\end{table}  

\subsection{$\it{B}$ phases}
\label{sec:Bphase}
\subsubsection{Odd number of sites}
\label{sec:Boddnumber}

In the $B_1$ sub-phase, the ground state has total spin $S^z=-\frac{1}{2}$   which corresponds to a static spin distribution and is represented by 
\be
|-\frac{1}{2}\rangle
\ee

Unlike in the $A$ phases, there exists only a single boundary bound state solution corresponding to the bound state at the left edge. Starting from the ground state, this bound state can be added (which has spin $S^z=\frac{1}{2}$) by adding a spinon whose spin orientation can be either in the positive or negative $z$ direction resulting in the state with total spin $S^z=\pm \frac{1}{2}$ respectively. This state has energy $E_0+m_L+E_{\theta}$, and hence has the lowest energy in the limit $\theta\rightarrow\infty$. It is represented by

\be
|\pm \frac{1}{2}\rangle_{L}
\ee

\vspace{1mm}

In the sub-phase $B_2$, the state which does not contain a bound state at either edge contains a spinon whose spin orientation is either in the positive or negative $z$ direction. The energy of this state is $E_0+E_{\theta}$ and thus forms a continuous branch parameterized by $\theta$. The ground state is obtained in the limit $\theta\rightarrow\infty$ and is represented by

\be
|\pm \frac{1}{2}\rangle
\ee

Starting from this ground state one can add a bound state at the left edge (which has spin $S^z=-\frac{1}{2}$) by removing the existing spinon. The resulting state has energy $E_0+m_L$ with total spin $S^z=-\frac{1}{2}$, and is represented by \be |-\frac{1}{2}\rangle_L.\ee

\vspace{1mm}

 By using the transformation $L\rightarrow R$, the states in the phases $B_8$ and $B_7$ can be obtained by starting with the states in the phases $B_1$ and $B_2$ respectively.  By using the transformation \ref{z2}, the states in the phases $B_5,B_6,B_3$ and $B_4$ can be obtained from the states in the phases $B_1,B_2,B_7$ and $B_8$ respectively.


\begin{table}[h!]
\centering
\caption{Energies and local bound state  parities of the ground state and the lowest energy states corresponding to each tower in all the $B$ phases for odd number of sites is shown below.}

\begin{tabular}{|c|c|ccc|}
\hline
\hline
\;\;Phase\;\;&\;\;State\;\;&\;\;Energy-$E_0$\;\;&\;\;$\mathcal{P}_L$\;\;&\;\;$\mathcal{P}_R$\;\;\\
\hline
$B_1$&$|-\frac{1}{2}\rangle$& 0 \;(g.s)& 1 & 1\\
\;&$|\pm\frac{1}{2}\rangle_{L}$& $m_L$ & -1 & 1\\

\hline

$B_8$&$|-\frac{1}{2}\rangle$& 0\;(g.s) & 1 & 1\\
\;&$|\pm\frac{1}{2}\rangle_{R}$&  $m_R$& 1 & -1\\

\hline

$B_2$&$|-\frac{1}{2}\rangle_{L}$& $m_L$& -1 & 1\\
\;&$|\pm\frac{1}{2}\rangle$& 0 \;(g.s) & 1 & -1\\

\hline

$B_7$&$|\pm\frac{1}{2}\rangle$& 0 \; (g.s) & 1 & 1\\
\;&$|-\frac{1}{2}\rangle_{R}$&  $m_R$& 1 & -1\\

\hline

$B_4$&$|\frac{1}{2}\rangle$& 0\;(g.s)& 1 & 1\\
\;&$|\pm\frac{1}{2}\rangle_{R}$& $m_R$ & 1 & -1\\

\hline

$B_5$&$|\frac{1}{2}\rangle_{L}$& 0\;(g.s)& 1 & 1\\
\;&$|\pm\frac{1}{2}\rangle_{L}$& $m_L$  & -1 & 1\\

\hline

$B_3$&$|\pm\frac{1}{2}\rangle$& 0 \; (g.s) & 1 & 1\\
\;&$|\frac{1}{2}\rangle_{R}$&  $m_R$& 1 & -1\\

\hline

$B_6$&$|\pm\frac{1}{2}\rangle$& 0 \;(g.s) & 1 & 1\\
\;&$|\frac{1}{2}\rangle_{L}$&  $m_L$& -1 & 1\\
\hline
\hline
\end{tabular}
\label{tableBodd}
\end{table}

\subsubsection{Even number of sites}
\label{sec:Bevennumber}

In the phase $B_1$, the state with no bound states at both the edges is two fold degenerate. It contains a spinon on top of the static spin distribution of the ground state in the phase $B_1$ corresponding to odd number of sites case. The spin orientation of the spinon can be either in the positive or negative $z$ direction which results in a doubly degenerate state with total spin $S^z=0,-1$. This state has energy $E_0+E_{\theta}$ and thus forms a continuous branch which is parameterized by $\theta$. The ground state is obtained in the limit $\theta\rightarrow\infty$ and is represented by
\be |0\rangle,  |-1\rangle.\ee

We can add the bound state at the left edge (with spin $S^z=\frac{1}{2}$) to the ground state by removing the existing spinon. This results in a state 

\be
|0\rangle_L
\ee
with total spin $S^z=0$ with energy $E_0+m_{L}$.

\vspace{1mm}

In the phase $B_2$, the state which does not contain bound state at either edge has total spin $S^z=0$ and has energy $E_0$. It is represented by 

\be |0\rangle.  \ee

We can add the bound state at the left edge (with spin $S^z=-\frac{1}{2}$) by adding a spinon with spin oriented either in the positive or negative z-direction and hence resulting in a doubly degenerate state with total spin $S^z=-1,0$. This state has energy $E_0+E_{\theta}+m_L$, and hence the lowest energy of this state corresponds to the limit $\theta\rightarrow\infty$ and is represented by

\be
|0\rangle_L,  |-1\rangle_L
\ee

\vspace{1mm}

Similar to the odd number of sites case, the states in the phases $B_8$ and $B_7$ can be obtained by starting with the states in $B_1$ and $B_2$ respectively, by making the transformation $L\rightarrow R$. By using the transformation \ref{z2}, the states in the phases $B_5,B_6,B_3$ and $B_4$ can be obtained from the states in the phases $B_1,B_2,B_7$ and $B_8$ respectively.

 
 \vspace{1mm}

\begin{table}[h!]
\centering
\caption{Energies and local bound state parities of the ground state and the lowest energy states corresponding to each tower in all the $B$ phases for even number of sites is shown below.}

\begin{tabular}{|c|c|ccc|}
\hline
\hline
\;\;Phase\;\;&\;\;State\;\;&\;\;Energy-$E_0$\;\;&\;\;$\mathcal{P}_L$\;\;&\;\;$\mathcal{P}_R$\;\;\\
\hline
$B_1$&$|-1\rangle$, \;$|0\rangle$& 0 \;(g.s)& 1 & 1\\
\;&$|0\rangle_{L}$& $m_L$ & -1 & 1\\

\hline

$B_8$&$|-1\rangle$, \;$|0\rangle$& 0 \;(g.s)& 1 & 1\\
\;&$|0\rangle_{R}$& $m_R$ & 1 & -1\\

\hline

$B_2$&$|-1\rangle_{L}$,\; $|0\rangle_{L}$& $m_L$& -1 & 1\\
\;&$|0\rangle$& 0  \;(g.s) & 1 & 1\\

\hline

$B_7$&$|-1\rangle_{R}$,\; $|0\rangle_{R}$& $m_R$& 1 & -1\\
\;&$|0\rangle$& 0 \;(g.s) & 1 & 1\\
\hline

$B_4$&$|1\rangle$, \;$|0\rangle$& 0 \;(g.s)& 1 & 1\\
\;&$|0\rangle_{R}$& $m_R$ & 1 & -1\\

\hline

$B_5$&$|1\rangle$, \;$|0\rangle$& 0 \;(g.s)& 1 & 1\\
\;&$|0\rangle_{L}$& $m_L$ & -1 & 1\\

\hline

$B_3$&$|1\rangle_{R}$,\; $|0\rangle_{R}$& $m_R$& 1 & -1\\
\;&$|0\rangle$& 0 \;(g.s) & 1 & 1\\

\hline

$B_6$&$|1\rangle_{L}$,\; $|0\rangle_{L}$& $m_L$& -1 & 1\\
\;&$|0\rangle$& 0 \;(g.s) & 1 & 1\\

\hline
\hline
\end{tabular}
\label{tableBeven}
\end{table}  

Unlike in the $A$ sub-phases where there exists bound states at both the edges, we have seen that in $B$ sub-phases there exists only one bound state at either the left or the right edge. Similar to the $A$ sub-phases, excitations can be built up starting from the ground state and from the state containing a bound state either at the left or the right edge by adding even number of spinons, strings and quartets. This leads to the Hilbert space in each $B$ sub-phase consisting of only two towers. For example, in the phase $B_1$, the two towers have the bound state parities $\mathcal{P}_L=1,\mathcal{P}_R=1$ and $\mathcal{P}_L=-1,\mathcal{P}_R=1$, whereas in the $B_8$ phase they correspond to $\mathcal{P}_L=1,\mathcal{P}_R=1$ and $\mathcal{P}_L=1,\mathcal{P}_R=-1$.
The ground states and the lowest energy states corresponding to the two towers in all the $B$ sub-phases are summarized in the tables \ref{tableBodd}, \ref{tableBeven}.

\subsection{$\it{C}$ sub-phases}
\label{sec:Cphase}
\paragraph{Odd number of sites}
 In the sub-phases $\it{C_1}, \it{C_3}$, the ground state is represented by 
 \be |\mp\frac{1}{2}\rangle\ee 
 
 and have total spin $S^z=\mp\frac{1}{2}$ respectively, which corresponds to a static spin distribution. In the sub-phases $\it{C_2}, \it{C_4}$ the lowest energy state contains a spinon with spin pointing either in the positive or negative z-direction resulting in a two fold degenerate state parameterized by rapidity $\theta$. The ground state is obtained in the limit $\theta\rightarrow\infty$. The spin orientation of the spinon dictates the total spin $S^z=\pm\frac{1}{2}$ of the state. They are represented by \be|\pm\frac{1}{2}\rangle.\ee

 \paragraph{Even number of sites}
 In the sub-phase $\it{C_1}$, the lowest energy state contains a spinon with rapidity $\theta$ with spin oriented either in the positive or negative z-direction on top of the static spin distribution of the ground state in the sub-phase $C_1$ corresponding to odd number of sites case. This state is two fold degenerate and is parameterized by rapidity $\theta$. The ground state is obtained in the limit $\theta\rightarrow\infty$ and is represented by
 
 \be|0\rangle,  |-1\rangle \ee
 
 with total spin $S^z=0$, $S^z=-1$ corresponding to the spin orientation of the spinon which is along the positive and negative $z$ direction respectively. Similarly, in the sub-phase $\it{C_3}$, the lowest energy state contains a spinon with spin pointed either in the positive or negative z-direction with rapidity $\theta$ on top of the static spin distribution of the ground state in the phase $C_3$ corresponding to odd number of sites case.  It is two fold degenerate and is parameterized by rapidity $\theta$. The ground state is obtained in the limit $\theta\rightarrow\infty$ and is represented by
 
 \be|0\rangle,  |1\rangle\ee 
 
 with total spin $S^z=0$, $S^z=1$ corresponding to the spin orientation of the spinon which is along the negative and positive $z$ direction respectively. In the sub-phases $C_2,C_4$, the ground state has total spin $S^z=0$ and is represented by 
 
 \be |0\rangle. \ee 
 
 Similar to the $A$ and $B$ sub-phases, in each $C$ subphase, excitations can be built on top of the ground state by adding even number of spinons, strings and quartets generating a single tower of excited states which can be labelled by $\mathcal{P}_L=1,\mathcal{P}_R=1$.

\subsection{Eigenstate phase transition}
\label{sec:eigenstatept}
After this rather lenghtly, but complete, description of the excited states let us now summarize our results.
As we saw there exists a critical value of the edge fields $h_c$, $|h_c|=2$, at each edge associated with the existence of an edge bound state. When $|h_{i=(L,R)}| > 2$ a localized bound state is stabilized close to the corresponding edge  $i=(L,R)$. The three types of phases $A_j$, $B_j$ and $C_j$ distinguish themselves by the number of bound states  they support, i.e: two, one and zero. Independently of the parity of $N$ we showed that in the $A$-type phases the Hilbert space splits into four towers of excited states while there exists two towers in the $B$-type phases and only one tower in the $C$-type phases.
When compared to the ground state phase diagrams (see Figs.(\ref{fig:evengs},\ref{fig:oddgs})) each quadrant splits into one $C_j$ sub-phase, two $B_j$ sub-phases and one $A_j$ sub-phase as displayed in  the Fig. \ref{fig:PD}. At this point a natural question arises: what is the nature of the transition that occurs as one moves from an $A_j$ sub-phase to a $B_j$ sub-phase or from a $B_j$ sub-phase to a  $C_j$ sub-phase by varying the edge fields. 

Without loss of generality let us fix on quadrant with $h_L >0$ and $h_R > 0$. Consider first the situation where 
both $h_{L,(R)} >2$, that is one sits in the $A_1$  sub-phase. Then let  the left boundary magnetic field $h_L$  be fixed while the right boundary fields $h_R$ is decreased. As $h_R$  is decreased below the critical value $h_c=2$, we move into the sub-phase ${B}_1$. The two states which contain the bound state at the right edge no longer exist. On the boundary between the $A_1$ and $B_1$ sub-phases, the energy of the bound state and energy of the spinon with zero rapidity coincide $m_R\sim 2\pi=E_{\theta\rightarrow 0}$. Hence it is natural to interpret that the bound state at the right edge leaks into the bulk by taking the form of a spinon with rapidity $\theta \sim 0$. Similarly, moving from ${A}_1$ to ${B}_8$ (see Fig. \ref{fig:PD}), the bound state corresponding to left boundary leaks into the bulk.  Similarly, moving from ${B}_1$ to ${C}_1$, the value of the left boundary field takes values lesser than critical value, and hence the bound state present at the left edge leaks into the bulk in a similar way, resulting in $C_1$ having no bound states at either edge. The same phenomena of bound states leaking into the bulk occurs as one moves from any $A$ sub-phase into the respective $B$ and $C$ sub-phases. 

\begin{center}
\begin{figure}[!h]
\includegraphics[width=0.9\columnwidth]{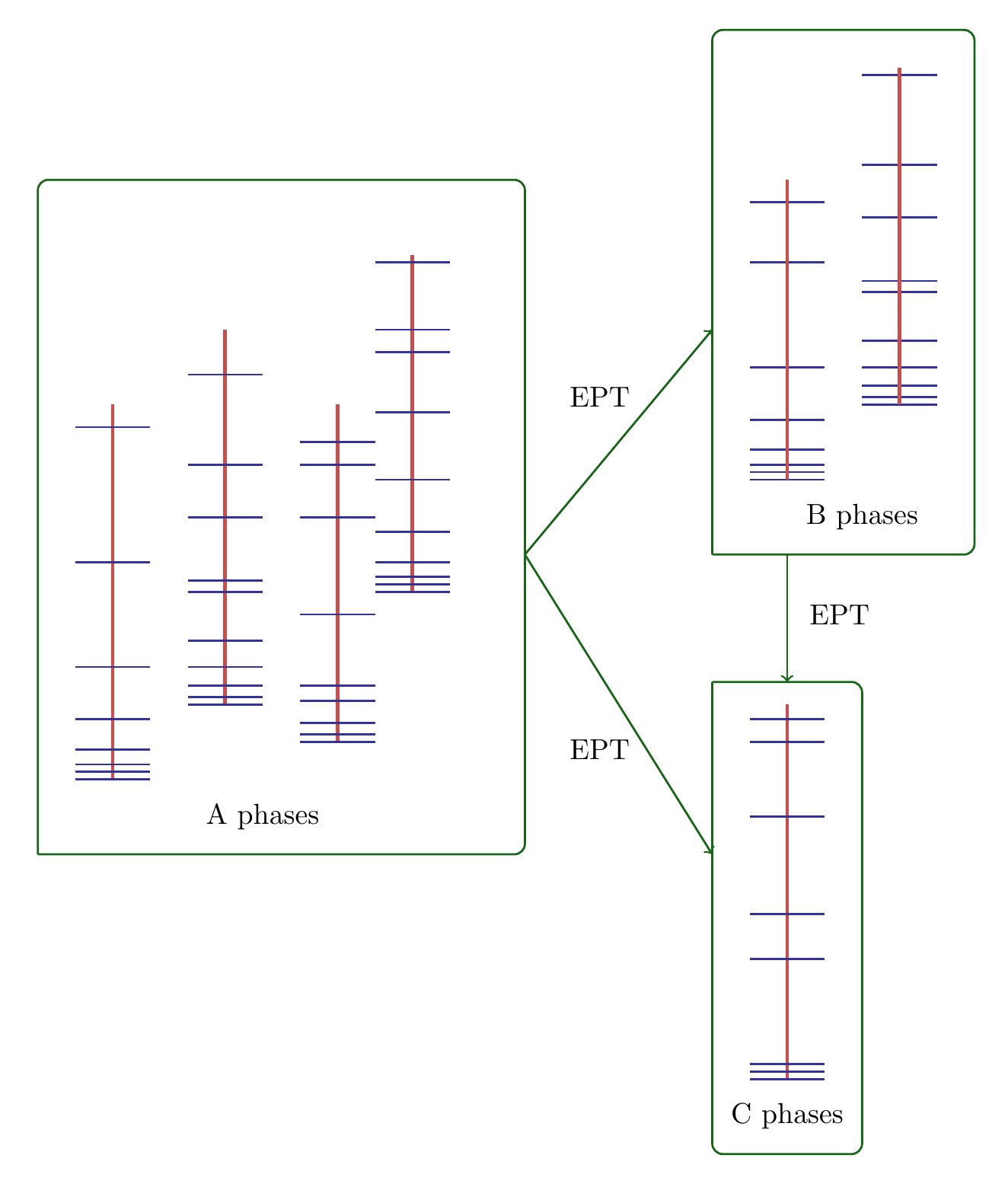}
\caption{The Hilbert space in the $A$ phases is comprised of four towers whereas it is comprised of two towers in the $B$ phases and a single tower in the $C$ phases. Figure illustrates the eigenstate phase transitions (EPT) that occur between $A$, $B$ and $C$ type phases, where the number of towers of the Hilbert space changes. 
\label{fig:EPT}}
\end{figure}
\end{center}

More importantly, associated with the appearance or disappearance of localized bound states is the fact that when one goes from any sub-phase to another, the whole structure of the Hilbert space changes. The excited states organize themselves into towers whose number is different in the $A$, $B$ or $C$ type phases. We saw that the towers are labelled by  additional quantum numbers which are the bound state parities $ \mathcal{P}_{L,R}$ (see Eq.(\ref{bsparity})). The four towers in $A$-type phases are labeled by $(\mathcal{P}_{L}, \mathcal{P}_{R})=(\pm 1, \pm 1)$, the two towers in the $B$-type phases
by $(\mathcal{P}_{L}, \mathcal{P}_{R})=(\pm 1, +1)$ and $(\mathcal{P}_{L}, \mathcal{P}_{R})=(+1, \pm 1)$ and the unique tower of the $C$-type phases by 
$(\mathcal{P}_{L}, \mathcal{P}_{R})=(+1, +1)$. It is interesting to notice that 
one may have also labeled these towers in the $A$-type phases by the spinon parity
$\mathcal{P}_{s}=(-1)^n$, where $n$ denotes the number of spinons in a given eigenstate of the Hamiltonian. The relation between $\mathcal{P}_{L}, \mathcal{P}_{R}$ and $\mathcal{P}_{s}$ depends on the phase and the parity of the number of sites $N$ as follows
\bea
\mathcal{P}_{L}\mathcal{P}_{R} (-1)^n (-1)^N =(-1)^k,
\eea
 where $k$ labels the different sub-phases $A_k$ as given in Fig.(\ref{fig:PD}). As when crossing from an $A_j$ sub-phase to either a $B_j$ or $C_j$ sub-phase the structure of the Hilbert space changes, and we coin the corresponding phase transition a (boundary) eigenstate phase transition. Such transitions might be probed through dynamical properties at infinite temperatures that involve operators localized close enough to the boundaries. We shall elaborate  on this topic in a forthcomming work.
 At present we shall content ourselves, in the next section, with the  simpler question of how this transition reflects itself in  the ground state properties of local observables.

\section{Ground state magnetization profile and spin accumulation}  
\label{sec:numerical}

\begin{figure}[t]
    \centering
    \includegraphics[width=0.49\linewidth]{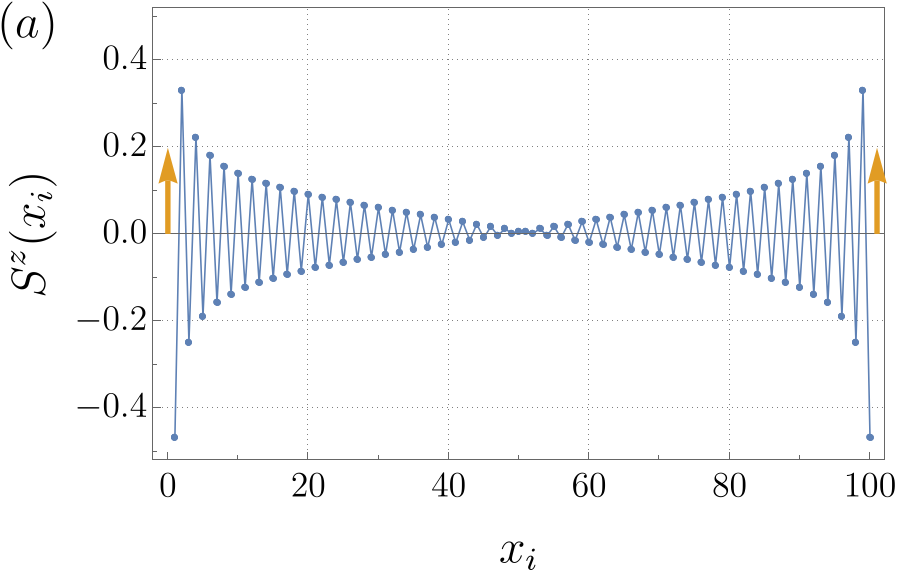}
    \includegraphics[width=0.49\linewidth]{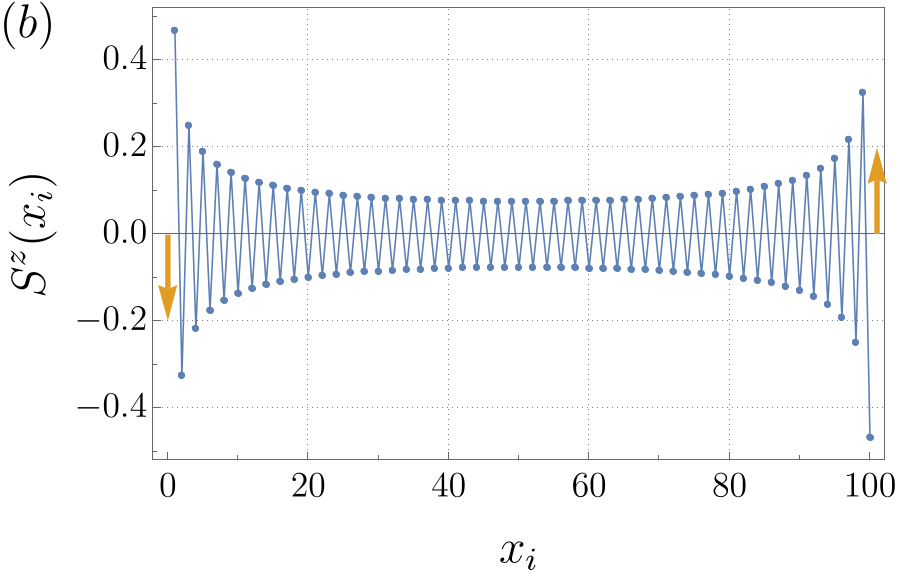}
    \includegraphics[width=0.49\linewidth]{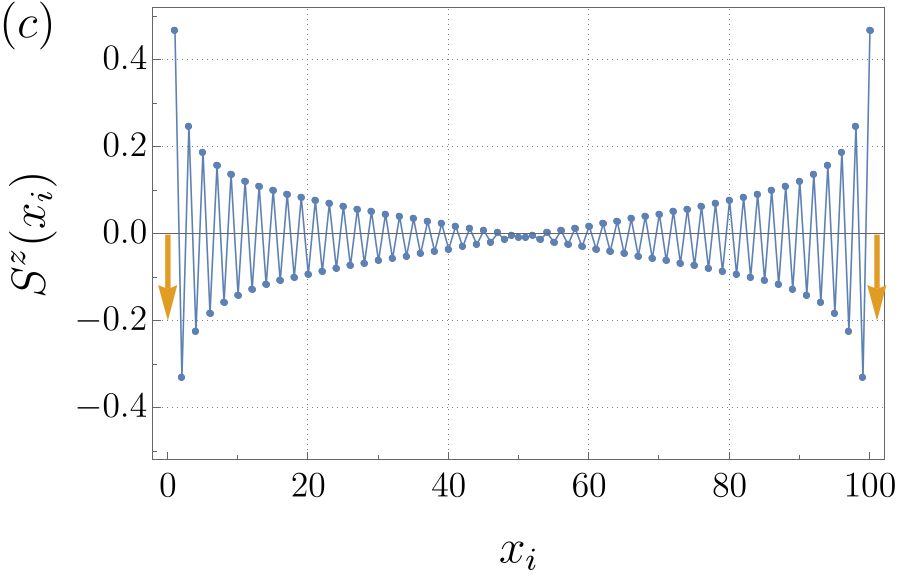}
    \includegraphics[width=0.49\linewidth]{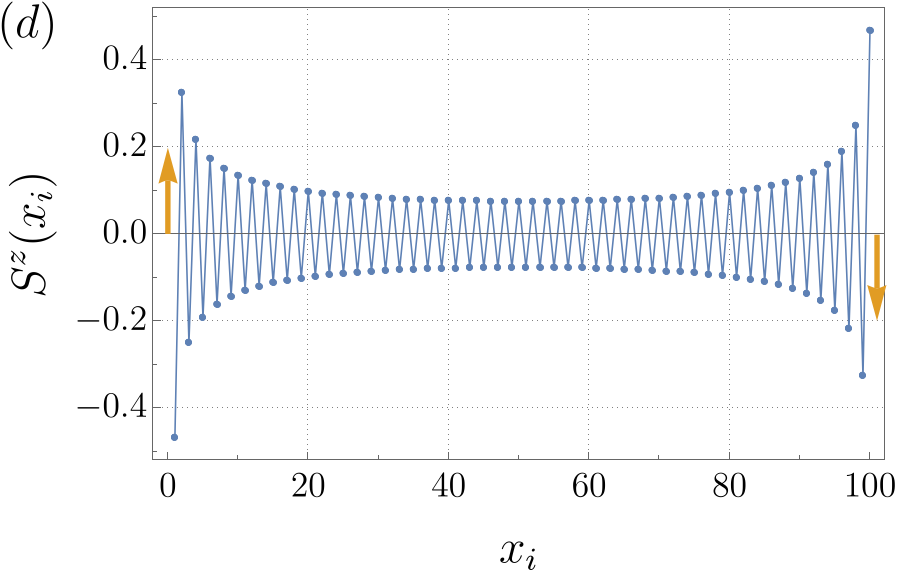}
    \caption{The magnetization profile of the ground state in (a) $A1$, (b) $A2$, (c) $A3$, and (d) $A4$ phase, respectively. The boundary fields are $h_L = \pm 4$, $h_R = \pm 4$ on a chain of length $N=100$, and the arrow indicates the direction of the boundary fields.
    }
    \label{fig:Sz}
\end{figure}  

To this end we shall be interested in the behavior of the {\it local} magnetization  profile  induced by the edge magnetic fields  in the different sub-phases.  We shall  use the DMRG method, which is ideal for one-dimensional systems, to calculate the ground state of Eq.~\eqref{eq:H} on finite size systems. In particular, we considered system sizes up to $N=1600$ sites, where a maximum of $1000$ states are kept to keep the truncation error below $10^{-12}$. The DMRG calculations in this paper are performed using the ITensor Library~\cite{itensor}.
Once the ground state is obtained we  compute the spin expectation value $S^z (x_i)\equiv \langle \sigma_i^z /2 \rangle$ as a function of position $x_i$ for various 
edge magnetic fields $h_L,h_R$ in the different phases of the problem.

\subsection{Magnetization Profile}
\label{sec:corr}

From the magnetization obtained from the DMRG calculations, we use the following ansatz for the magnetization near the boundary:
\begin{align}
    S^z (x_i) = (-1)^{i} \left( A + \frac{B}{\sqrt x_i} + C e^{-x_i/\xi} \right) + \frac{D}{x_i}.
    \label{eq:fit}
\end{align}
 We have introduced a constant staggered magnetization ($A$) that vanishes in the thermodynamic limit as well as the 
alternating $1/\sqrt{x}$  and  uniform  $1/x$ terms
which account, to leading order, for the gapless  bulk. 
In the  bosonization language they correspond to the staggered and uniform component of the magnetization in the long distance limit.  In addition to the above terms, we have also included a term which goes like $\sim e^{-x_i/\xi}$ to account for any exponentially localized spin accumulation.

Overall we find excellent agreement between our  DMRG results and its fit (\ref{eq:fit}) for several values of $h_L,h_R$. We show in Fig.~\ref{fig:Sz}, as an example, the magnetization profile $S^z (x_i) $ deep in  the four $A$-phases for boundary fields $|h_L|=|h_R|=4$ and system size  $N=100$. As expected, the magnetizations at the boundaries are all opposite to the boundary field directions and one clearly observes a spin accumulation close to the edges. We notice though that, since  the bulk is gapless, the spin accumulation is not expected to be sharply localized at the edge. This can be seen in the magnetization profile which exhibit asymptotic  power-law antiferromagnetic decay  sufficiently far away from the edges. Notice that, since in  the $A_1$ and $A_3$ phases the ground state  contains a spinon in the bulk for even chains,  a node in the bulk antiferromagnetic configuration [Fig.~\ref{fig:Sz}(a)(c)] is clearly seen. Alternatively, for odd chain lengths the $A_1$ and $A_3$ phases  have no spinons whereas the $A_2$ and $A_4$ phases do.


\begin{figure*}
    \centering
    \includegraphics[width=0.31\linewidth]{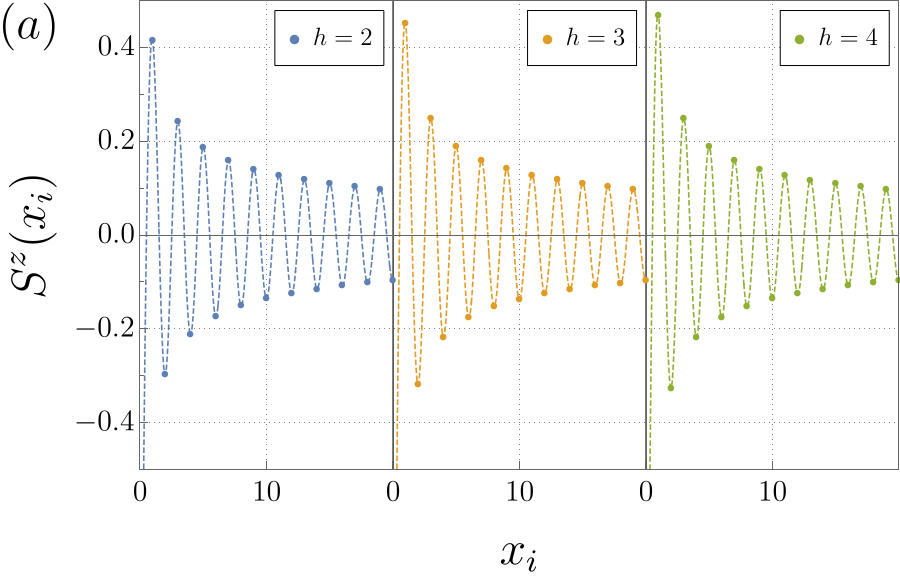}
    \includegraphics[width=0.36\linewidth]{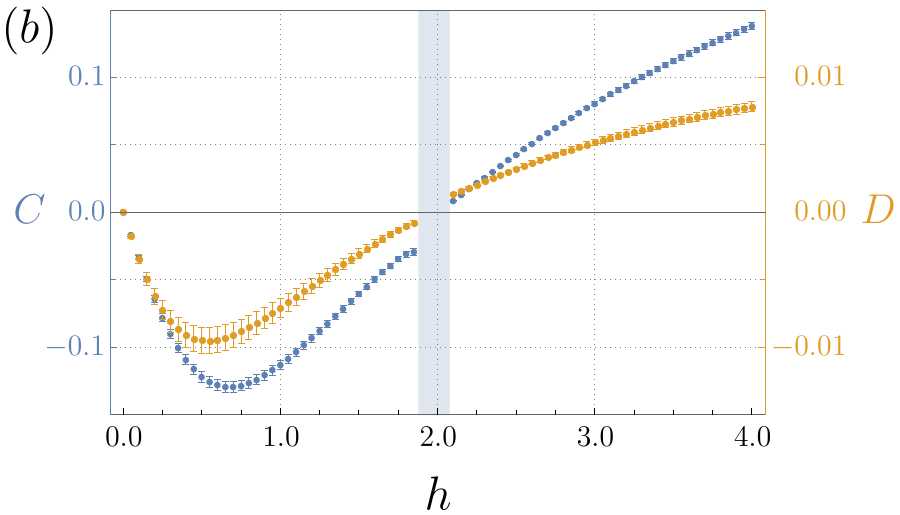}

   \caption{(a) Magnetization (dots) and its fit to Eq.~\eqref{eq:fit} (dashed lines) for the first 20 sites of $N=1000$ chain. (b) The fit-parameters $C$ and $D$ from Eq.~\eqref{eq:fit}, as a function of $h$. Both parameters vanishes at the critical $h_c=2$. The shaded region is where the fitting to Eq.~\eqref{eq:fit} numerically fails as it is close to $h_c=2$. (See Appendix~\ref{sec:fitting} for more details)
    }
    \label{fig:xi1}
\end{figure*}

\begin{figure*}
    \centering
    
    \includegraphics[width=0.28\linewidth]{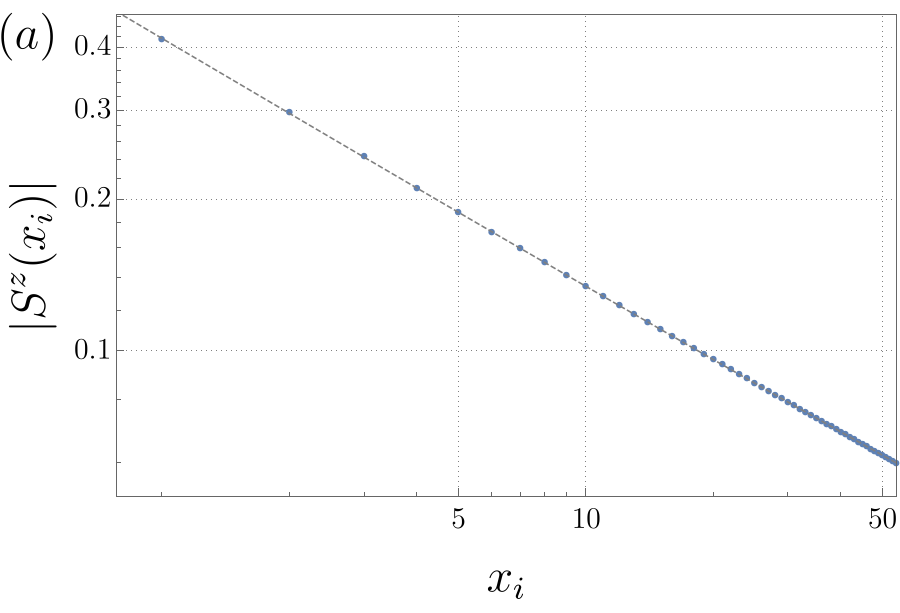}
    \includegraphics[width=0.335\linewidth]{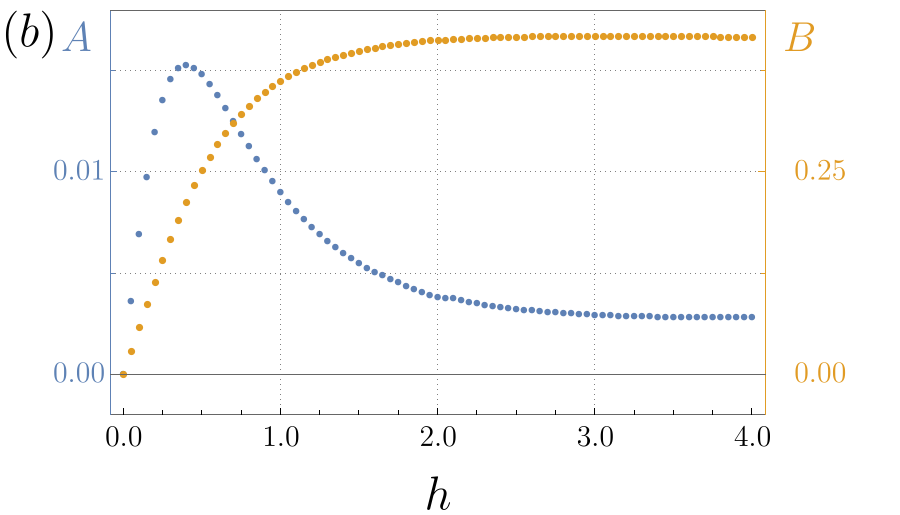}
    \includegraphics[width=0.286\linewidth]{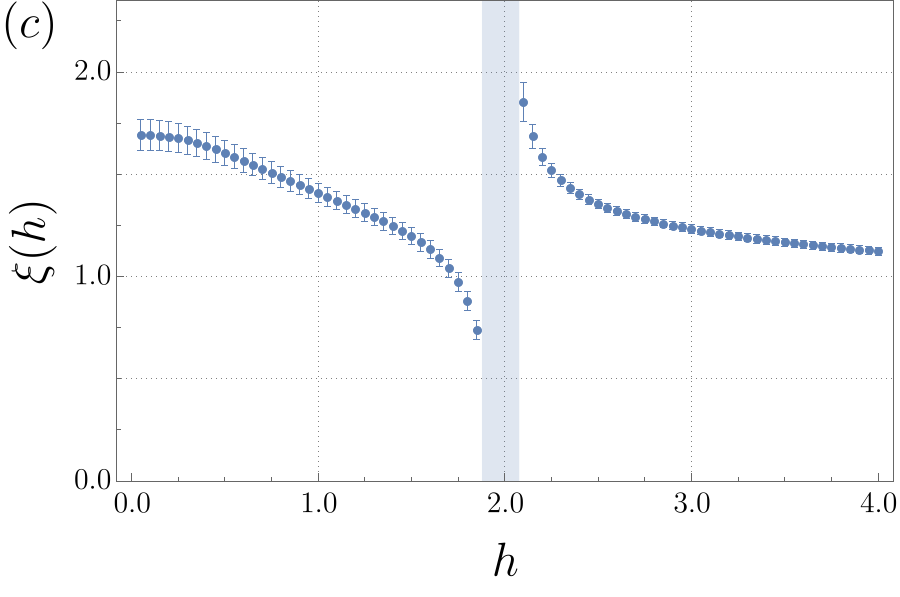}
    
   \caption{(a) The magnetization for the $N=1000$ chain with critical boundary fields ($h = h_c = 2$). The magnetization of a critical chain decays as $\sim 1/\sqrt{x_i}$ as can be seen from the good agreement with the $|S^z(x_i)| = 0.413/\sqrt{x_i} + 0.00385$ obtained from fitting the data. (b) The fit-parameters $A$ and $B$ from Eq.~\eqref{eq:fit}, as a function of boundary field $h$. (c) The length scale $\xi$ of the exponentially localized boundary spin, also from Eq.~\eqref{eq:fit}, as a function of $h$. The critical point is $h_c=2$ where the length scale diverges. 
    }
    \label{fig:xi2}
\end{figure*}

We now discuss in more detail about our results for the fit in the particular case $h_L=h_R=h$, where the system exhibits a $\mathbb{Z}_2$ space parity symmetry.
As one varies $h$, this allows us to study the spin magnetization profile when going from the $C_1$ phase to the $A_1$ phase (see Fig \ref{fig:PD}).
We show for example in Fig. \ref{fig:xi1}(a) the magnetization profile for different values of $h$ in the $A_1$ phase and the critical point. 
Fitting the DMRG data with the form (\ref{eq:fit}), we can extract the parameters $A$, $B$, $C$ and  $D$ as well as  the length scale  $\xi$ as a function of $h$. 
 The coefficients $C$ of the exponential term and $D$ of the uniform component $1/x$ are shown in Fig.~\ref{fig:xi1}(b).  The constant term $A$ and the coefficient $B$ of the staggered magnetization component $1/\sqrt{x}$ are shown in Fig.~\ref{fig:xi2}(b). Finally   $\xi$ is shown in Fig.~\ref{fig:xi2}(c). 

 From our data we first observe that when $h=h_c=2$ the magnetization profile takes a particular simple form
 in the thermodynamical limit
\be
S^z(x_i) \simeq (-1)^i 0.413/\sqrt{x_i}, 
\label{magh=2}
\ee
as both $C$ and $D$ are zero (even at a finite size)  and $A$ goes to zero as $N\rightarrow \infty$. We display in  Fig. \ref{fig:xi2}(a) our best fit for the magnetization when $h=h_c=2$ and $N=1000$ which shows an almost perfect 
$\sim 1/\sqrt{x}$ behavior asymptotically. It is remarkable that for this value of the edge field the bulk uniform component $1/x$ disappears from the magnetization profile. 

This case $h=h_c=2$ seems to play a special role in the magnetization profile. Indeed we find that both the coefficient $C$ and $D$ change sign when going accross the $h=h_c=2$ point where they vanish. The change in the sign of $C$ means that the exponential term enhances (diminishes) its contribution to  $S^z(x_i)$ when $h>h_c$ ($h<h_c$). 
On the other hand the change of sign of $D$ can be interpreted as a $\pi$ phase-shift of the uniform component term in Eq.~\eqref{eq:fit}. At the transition, both contributions vanish as we find $C,D=0$. Another important feature is that coefficients $A$ and $B$ saturate as one increases $h$ above $h_c=2$ as seen in Fig.~\ref{fig:xi2}(b). This means that at magnetic fields $h$  larger than $h_c=2$ the constant contribution as well as the  staggered component of the magnetization are insensitive to the  edge magnetic field. As these are the dominant contributions for large $x$, this means that 
the magnetizations far from the edges are essentially insensitive to the edge fields when $h > h_c$, in contrast to low fields $h < h_c$ where both $A$ and $B$ wildly varies.

In the light of the discussion given in  the precedent section it seems that at the critical point where the eigenstate phase transition occurs (here between the $A_1$ and $C_1$ phases on the $\mathbb{Z}_2$ symmetric line $h_L=h_R$)  a qualitative change also occurs in the magnetization profile in the ground state. Whether this change corresponds to a genuine critical point for the ground state properties (reflecting itself into a singular behavior of  the magnetization profile at $h_c$)  is a non trivial issue. Indeed, although we find that the 
length scale $\xi$ diverges as $h$ approaches $h_c$, at the same time $C,D\rightarrow 0$ and the numerical fitting overfits the data in the vicinity of the critical point ($h=h_c=2$).
We mark this region with a grey shade in Fig.~\ref{fig:xi1} (b) and Fig.~\ref{fig:xi2} (c) (for more detail see Appendix~\ref{sec:fitting}). We therefore find it difficult to conclude that  these coefficients, or length scale $\xi$,  serve as  order parameters for a genuine ground state phase  transition.  

\subsection{Boundary spin accumulation}

\begin{figure*}
    \centering
    \includegraphics[width=0.33\linewidth]{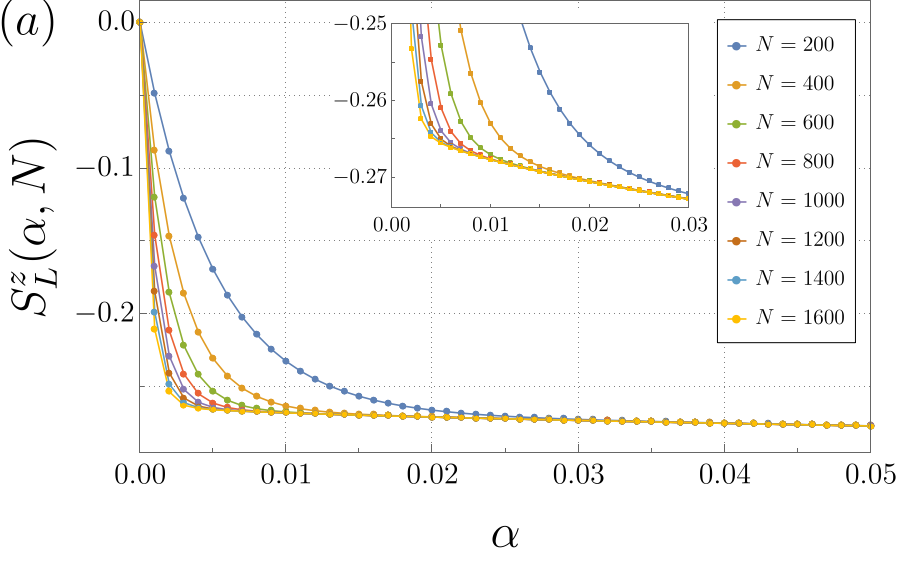}
    \includegraphics[width=0.33\linewidth]{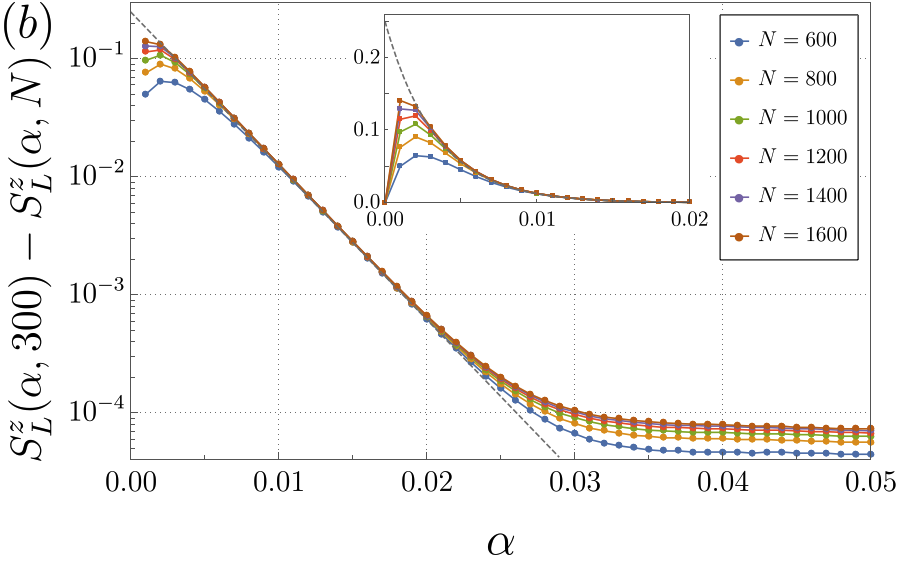}
    \includegraphics[width=0.31\linewidth]{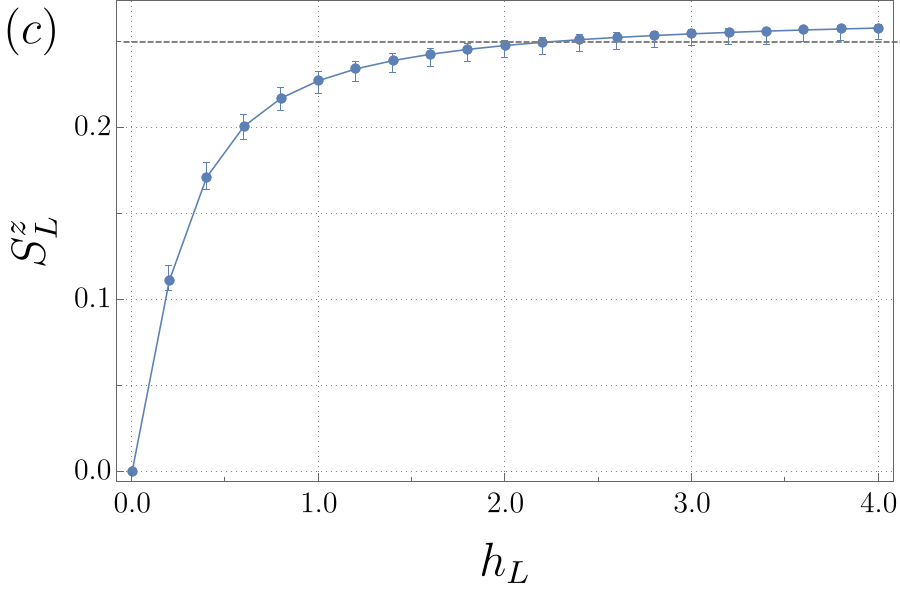}
    \caption{
    (a) The spin accumulation $S^z_L(\alpha, N)$ for various system sizes when $h_L = -h_R = 4$. The inset is a blowup near $\alpha = 0$. 
    (b) The difference between the spin accumulation for system size $N$ and that of the system of $N=300$. The inset is the same plot in linear scale, and the gray dashed lines are the conjectured $(1/4)e^{-300 \alpha}$. (c) The boundary spin accumulation obtained from the linear fit of (b) to $A e^{-300 \alpha}$ as a function of boundary field $h_L$. 
    }
    \label{fig:spin1/4}
\end{figure*}    

Due to the edge magnetic fields we naturally expect that some amount of spin is getting accumulated close to the edges. 
To calculate the spin accumulation associated with the edges of the system, we use the following definition of the spin accumulation on the left boundary $S^z_L$ as~\cite{Jackiw}:
\begin{align}
    S^z_L = \lim_{\alpha\rightarrow0} \lim_{N\rightarrow\infty} S^z_L(\alpha,N),
    \label{eq:szl}
\end{align}
where,
\begin{align}
    S^z_L(\alpha,N) = \sum_i e^{-x_i \alpha} S^z(x_i).
\end{align}
Note the order of the limit is relevant and it is important to take the thermodynamic limit first. 
Otherwise, the $\alpha \rightarrow 0$ limit removes the cutoff and the result becomes merely the total $S^z$ of the system.
Also, $\alpha \gg 1/L$ should be satisfied for the cutoff to be meaningful. 
In the following we shall compute $S^z_L$ in both the $C_2$ and $A_2$ phases
where $h_L=-h_R=-h$. The reason for this is to keep the odd parity of the system which simplifies the finite-size conjectures which will follow.

To obtain $S^z_L$  systematically, we infer $\lim_{N\rightarrow\infty} S^z_L(\alpha,N) \equiv S^z_L(\alpha)$ from the finite size calculations.
Fig.~\ref{fig:spin1/4}(a) shows the $S^z_L(\alpha,N)$ as a function of $\alpha$ for different system sizes, when $h = 4$. 
One important observation is that, as the system size grows $S^z_L(\alpha,N)$ converges to the $S^z_L(\alpha)$ curve. 
Therefore, each $S^z_L(\alpha,N)$ is converged to $S^z_L(\alpha)$ for $\alpha$ larger than a certain value of $\alpha_N$, which $\lim_{N\rightarrow\infty} \alpha_N = 0$. 
We conjecture the leading difference of the finite $S^z_L(\alpha,N)$ and infinite $S^z_L(\alpha)$ as:
\begin{align}
    S^z_L(\alpha) = S^z_L(\alpha,N) -\frac{1}{4} e^{-N \alpha}+\cdots,
   \label{eq:1/4}
\end{align}
for $h \geq h_c$.
This equation is suggestive that it consists of the proposed spin (1/4) and the only length scale ($N$), and directly implies the fractional 1/4-spin.
Taking the $\alpha \rightarrow 0$ limit to Eq.~\eqref{eq:1/4}, the first term vanishes because $S^z_L (0, N) = \sum_i S^z (x_i)$ for any finite $N$ and the total $S^z$ is zero for the even chain in $A_2$ phase. 
The remaining terms give the fractionalized $S^z_L = \pm 1/4$ per Eq.~\eqref{eq:szl} where the sign depends on the direction of the boundary field.


To see how this work let us give an example of our finite size scaling procedure in the particular case $h=4$.
In Fig.~\ref{fig:spin1/4}(b) we plot $S^z_L(\alpha,300) - S^z_L(\alpha,N)$ for various $N$ for the same parameters in Fig.~\ref{fig:spin1/4}(a). 
As we expect from our conjecture [Eq.~\eqref{eq:1/4}], the difference converges to $\frac{1}{4} e^{-N \alpha}$ (dashed line) for large $N$. 
To quantify the numerical value of $S_L^z$, we find the best exponential fit to the plots similar to Fig.~\ref{fig:spin1/4}(b) for different $h$ values, and obtain the spin accumulation $S_L^z(h)$ as the overall coefficient.

Our final result for the spin accumulation $S_L^z$ as a function of $h$ is shown in Fig.~\ref{fig:spin1/4}(c). Our results for $S^z_L$  are  consistent with $S^z_L = 1/4$ for $h\geq h_c = 2$ and decreases for smaller values of $h$. 
We thus find that at large fields  $h\geq h_c = 2$ a {\it fractional} quarter spin is likely to be accumulated at the edge in the $A_2$ phase. The situation at hand is similar to what happens in topological one dimensional gapless Spin Triplet Superconductors (STS) where there also a  {\it fractional} spin-$1/4$ is  getting localized at the edge. However in the present case we do not expect  $S^z_L$ to be a sharp quantum observable in contrast with the STS case where eigenvalues of $S^z_L$ label the different  edge states of the system. 
The reason for this stems from the absence of a gap in the bulk of the Heisenberg chain in contrast with  STS. The best way to check this is to compute the variance of the operator $S^z_L$ [Eq.~\eqref{eq:szl}] 

\bea
    \sigma^2_S &=& \lim_{\alpha\rightarrow0}\lim_{N\rightarrow \infty} \sigma^2_S (\alpha, N), \nonumber \\
    \sigma^2_S (\alpha, N)&=& 
    \langle (S^z_L(\alpha, N))^2 \rangle - \langle S^z_L(\alpha, N) \rangle^2.  \eea

\begin{figure*}
    \centering
    \includegraphics[width=0.325\linewidth]{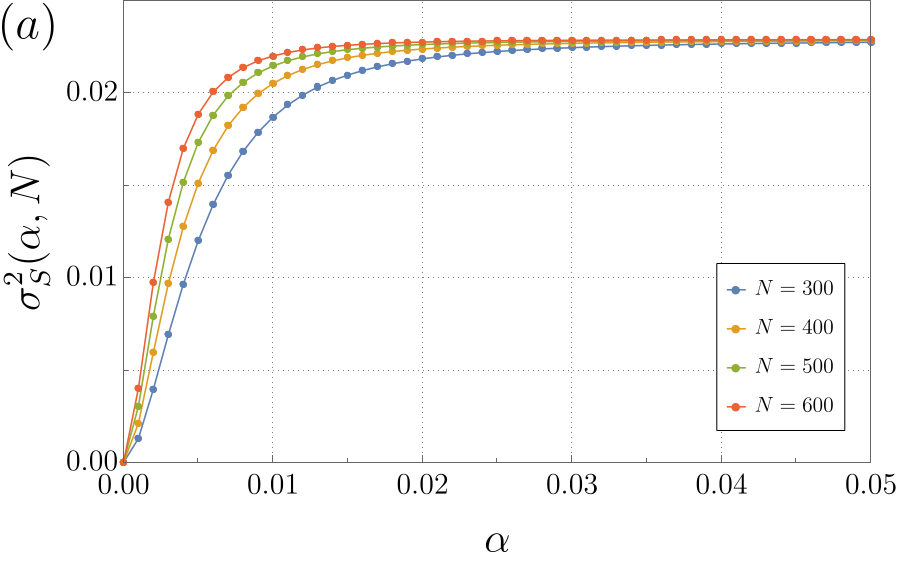}
    \includegraphics[width=0.32\linewidth]{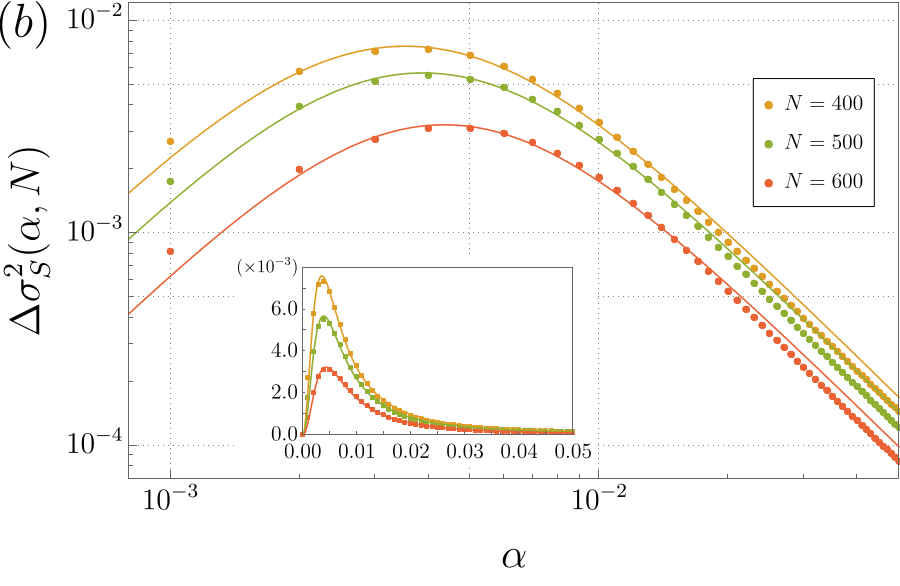}
    \includegraphics[width=0.325\linewidth]{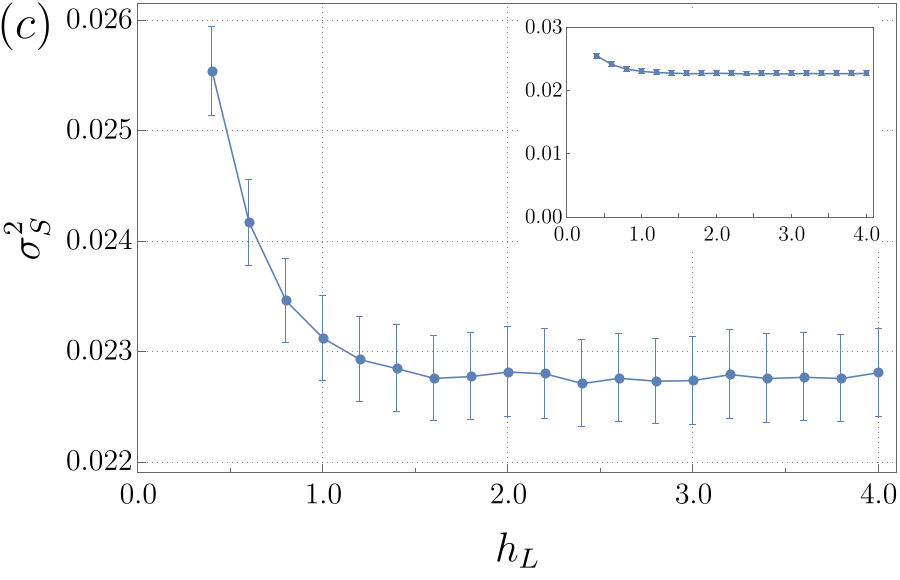}
    \caption{
    (a) The variance of $S^z_L (\alpha, N)$ for various system sizes when $h_L = - h_R = 4$. 
    (b) The difference between the variance for system size $N$ and that of the system of $N = 300$. The linear behavior for large $\alpha$ in this log-log plot suggests the difference is asymptotically $\sim \alpha^{-2}$. The solid lines are the result of a fit based on Eq.~\eqref{eq:var}. The discrepancies at large $\alpha$ are exaggerated due to the log scale. The inset is the same figure in linear scale which shows good agreement between data and fit.
    (c) The variance obtained by the fit parameters of (b). Variance remains nonzero beyond the phase transition. Inset is the same quantity plotted in a $y$-axis range down to zero.
    }
    \label{fig:var}
\end{figure*}

We show in Fig.~\ref{fig:var}(a)  $\sigma^2_S (\alpha, N)$ for different system sizes and $h=4$.
We observe the $\sigma^2_S (\alpha, N)$ converges to $\sigma^2_S (\alpha)$ as system size increases. We are thus led to
conjecture the leading finite-size correction to the infinite $N$ limit:
\begin{align}
    \sigma^2_S (\alpha) = \sigma^2_S (\alpha, N) + \frac{a}{(N \alpha)^2 + b} + \cdots.
    \label{eq:var}
\end{align}
This conjecture is based on the empirical observation that the difference of two curves in Fig.~\ref{fig:var}(a) follows $1/\alpha^2$ for large $\alpha$ and remains finite at $\alpha = 0$, which is qualitatively  reminiscent of how the connected spin-spin correlation function (that the variance is related to) vanishes for large momentum (represented by $\alpha)$.
Fig.~\ref{fig:var}(b) shows $\Delta \sigma^2_S (\alpha, N) \equiv \sigma^2_S (\alpha, N) - \sigma^2_S (\alpha, 300)$, together with the fitted equations using Eq.~\eqref{eq:var}. 
Importantly, we find that $a$ and $b$ are essentially independent of $N$ (with discrepancies within $1.61\%$) demonstrating the quality of the conjectured functional form with only two fit parameters.

We again take the $\alpha \rightarrow 0$ limit of Eq.~\eqref{eq:var} and obtain $\sigma^2_S$. 
Using the fitted parameters $a$ and $b$ from $\sigma^2_S (\alpha, 600) - \sigma^2_S (\alpha, 300)$, we get $\sigma^2_S = \sigma^2_S(0,N) + a/b$.
Since $\sigma^2_S$ and $a/b$ does not depend on $N$, the remaining term $\sigma^2_S (0,N)$ should also be $N$ independent. 
We indeed find that $\sigma^2_S (0,N)$ is nearly zero, and three orders of magnitude smaller than $a/b$.
From this, we plot $\sigma^2_S (\approx a/b)$ as a function of $h$ in Fig.~\ref{fig:var}(c).
Although the variance decrease as $h$ increase and crosses the phase transition, it remains nonzero.
This means that $S^z_L$ 
does not represent a sharp  quantum observable,
and as such, we cannot, in contrast with topological STS  superconductor;  label the states according to the value 
of the spin accumulations at the edges.


Overall we find that the ground state properties, as probed by the magnetization profile, displays interesting and non trivial behavior as a function of the edge fields $h$. In particular we observe that above the critical value $h> h_c=2$ a fractional spin $1/4$ is likely to be accumulated at the edge of the chain. However, in contrast with what happens in topological spin triplet superconductor, this fractional spin is not a sharp quantum observable and cannot be used as a genuine quantum number.  The reason for this stems from the absence of a gap in the bulk.
Nevertheless, the behavior of the spin accumulation as well as the  change of the magnetization profile  in the ground state might reflect, at least qualitatively,     the eigenstate phase transition between phases $A$ and $C$ discussed in the previous section.

\section{Discussion }
\label{sec:discussion}
We considered spin $\frac{1}{2}$ Heisenberg chain with boundary magnetic fields and analyzed it analytically using Bethe ansatz and also numerically using DMRG. Although the Heisenberg chain
has been immensely studied and is very well understood, the results that we have presented in this work have not been found before. We find that the system exhibits four different ground states for four different orientations of the boundary fields. The total spin $S^z$ in each ground state may differ, which depends on the orientation of the boundary fields and also depends on the evenness or oddness of the number of sites of the chain. As the orientation of the magnetic fields is changed, the system undergoes a phase transition where the ground state of the system changes. The nature of this phase transition is currently unknown to us and will be analyzed in the future work.

For a given orientation of the boundary fields, the system exhibits a high energy bound state exponentially localized at an edge when the boundary magnetic field 
takes values greater than the critical field $h_c$. Every phase corresponding to a certain ground state can be further divided into four sub-phases.  In one of the sub-phases the system exhibits bound states at both edges, and in one sub-phase the system exhibits no bound states while in the remaining two sub-phases the system exhibits a bound state at the left or the right edges. 

Starting from the ground state, one can build up excitations in the bulk by adding spinons, strings, quartets etc., and one obtains a tower of excited states. Similarly, starting from either the state which contains one bound state at the left or the right edge or from the state which contains two bound states, one can build up excitations in the bulk and one obtains different towers of excited states.  Hence in the region where the system exhibits two bound states, the Hilbert space is comprised of four towers and in the regions where the system exhibits one bound state, the Hilbert space is comprised of two towers and in the regions where there exists no bound states, the Hilbert space is comprised of a single tower of excited states.

 For a particular orientation of the boundary fields where the system exhibits a certain ground state, as the values of the magnetic fields is changed, the system undergoes an eigenstate phase transition, where the system may gain or lose a bound state at a particular edge which results in the change in the number of towers in the Hilbert space. Across this phase transition line where the structure of the Hilbert space changes, the total spin $S^z$ of the ground state of the system remains unchanged. To analyze the properties of the ground state across this phase transition, we chose the $\mathbb{Z}_2$ symmetric point where the values of the magnetic fields at the edges take equal values. By using DMRG we obtained the edge magnetization profile in the regions where the magnetic fields take values greater than and lesser than the critical field $h_c$. We find that when both the magnetic fields take values greater than the critical field, the total spin accumulation at  each edge saturates to $\frac{1}{4}$. To check whether this fractional spin is a genuine quantum observable, we calculated the variance and found that although it saturates to a small value, it remains non zero, indicating that the fractional spin $\frac{1}{4}$ is not a genuine quantum number. Nevertheless, in the region where both the magnetic fields take values greater than the critical field, there exists non zero probability to observe a non zero spin $S^z$ close to each edge. 

Although there exist no genuine spin fractionalization in the ground state, the structure of the Hilbert space and the eigenstate phase transition the system exhibits are also found in the gapless superconductors which exhibit SPT. Even though the Hilbert space is comprised of towers of excited states, unlike the gapped regime of the XXZ spin $\frac{1}{2}$ chain which exhibits SSB, degenerate pairing in the spectrum is not apparent due to the gapless nature of the bulk excitations. Although the model we considered is integrable, the structure of its Hilbert space and the eigenstate phase transition it exhibits might provide insight into systems with disorder which exhibit phenomena such as many body localizaton.

\acknowledgements
 J.H.P. thanks Wolfgang Ketterle for useful discussions about realizing boundary fields in experiment. P.A and P.R.P thank 
 F. H. L. Essler for very helpful discussions. J.L. and J.H.P. are partially supported by   the Air Force Office of Scientific Research under Grant No.~FA9550-20-1-0136 and the Alfred P. Sloan Foundation through a Sloan Research Fellowship.
J.H.P. acknowledges the Aspen Center for Physics, where some of this work was discussed, which is supported by National Science Foundation grant PHY-1607611. P.R.P. acknowledges support from Rutgers HEERF Fellowship during the stay at Rutgers University, where most part of the work was done.

 \bibliography{xxxmag}

\begin{widetext}
\appendix

 \section{Solution of Bethe ansatz equations}
 \label{appendix1}
In this section we provide a detailed calculation of the ground state and the boundary excitations for both odd and even number of sites chain in the phases $A_1$, $B_1$ and $C_1$, and describe how the solution in all the other phases can be constructed using the solution obtained in these phases. 

\subsection{$A$ phases}

Consider the phase $A_1$. In this phase both the magnetic fields point in the positive $z$-direction and take values $h_L,h_R>h_C$, ($h_C=2$), which translates to the boundary parameters taking values $0<p_L,p_R<\frac{1}{2}$. The Bethe equations corresponding to Bethe reference state with all down spins are \cite{Sklyannin,ODBA}

\bea
\label{BaeAdown}
\left(\frac{\lambda_j-\frac{i}{2}}{\lambda_j+\frac{i}{2}}\right)^{2N}\left(\frac{\lambda_j+i(\frac{1}{2}+p_{L})}{\lambda_j-i(\frac{1}{2}+p_{L})}\right)\left(\frac{\lambda_j+i(\frac{1}{2}+p_{R})}{\lambda_j-i(\frac{1}{2}+p_{R})}\right)\nonumber=\prod_{j\neq k=1}^{M}\left(\frac{\lambda_j-\lambda_k-i}{\lambda_j-\lambda_k+i}\right)\left(\frac{\lambda_j+\lambda_k-i}{\lambda_j+\lambda_k+i}\right)\eea

The Eigenvalues of the Hamiltonian are given by 
 \bea 
\label{BaeEAdown}
 E=-\sum_{j=1}^{M}\frac{2}{\lambda_j^2+\frac{1}{4}}+N-1-\frac{1}{p_L}-\frac{1}{p_R}
 \eea

\subsubsection{$A_1$: Odd number of sites}

 Let us first consider a state with all real Bethe roots. By applying logarithm to \ref{BaeAdown} we obtain 

\bea
\label{LogbaeAdown}
(2N+1)\Theta(2\lambda_j)-\Theta\left(\frac{\lambda_j}{\frac{1}{2}+p_L}\right)-\Theta\left(\frac{\lambda_j}{\frac{1}{2}+p_R}\right)=\sum_{k\neq j, \sigma=\pm} \Theta(\lambda_j+\sigma\lambda_k)+\pi \nu(\lambda_j)\eea

 where $\Theta(x)=\text{ArcTan(x)}$. We introduced the counting function $\nu(\lambda)$, where $\nu(\lambda_j)=I_j$. Differentiating \ref{LogbaeAdown} and noting that $2\rho(\lambda)=\frac{d}{d\lambda}\nu(\lambda)$, we obtain

 \bea
 \label{IntAdownreal}
 (2N+1)a_{\frac{1}{2}}(\lambda)-a_{\frac{1}{2}+p_L}(\lambda)-a_{\frac{1}{2}+p_R}(\lambda)=\pi\delta(\lambda)+2\pi\rho_{|-\frac{1}{2}\rangle}(\lambda)+2\int\rho_{|-\frac{1}{2}\rangle}(\mu)a_1(\lambda-\mu)d\mu
 \eea

 where a hole at $\lambda=0$ is added as $\lambda=0$ is a trivial solution to the Bethe equations \ref{BaeAdown} and leads to a zero wavefunction \cite{ODBA}. The reason for the subscripts for the density distribution will become evident soon.
 
 Taking Fourier transform we obtain

\bea
 \label{denAdownreal}
 \tilde{\rho}_{|-\frac{1}{2}\rangle}(\omega)=\frac{(2N+1)e^{-\frac{|\omega|}{2}}-e^{-(\frac{1}{2}+p_L)|\omega|}-e^{-(\frac{1}{2}+p_R)|\omega|}-1}{2\left(1+e^{-|\omega|}\right)}
 \eea
  
 The total number of Bethe roots is $M_{|-\frac{1}{2}\rangle}= \tilde{\rho}_{|-\frac{1}{2}\rangle}(0)=\frac{N-1}{2}$, hence $S^z=-\frac{1}{2}$.  We represent this state by $|-\frac{1}{2}\rangle$. Notice that the number of roots is an integer only for a spin chain with odd number of sites. \smallskip

Using \ref{BaeEAdown}, one can find the energy of the states. Equation \ref{BaeEAdown} can be written as

\bea
\label{energyAdown}
E=-4\int a_{\frac{1}{2}}(\lambda)\rho(\lambda)+N-1-\frac{1}{p_L}-\frac{1}{p_R}
\eea

Using the density distribution \ref{denAdownreal} in the above equation we obtain

\bea
\label{energygroundAdown}
E_{|-\frac{1}{2}\rangle}=E_0=-(2N+1)\ln(4)+N-1+\pi+\sum_{i=L,R}\Psi\left(\frac{p_i}{2}\right)-\Psi\left(\frac{p_i-1}{2}\right)-\frac{1}{p_i}\eea

Where $\Psi$ is the digamma function. There exists two boundary string solutions $\lambda_{p_L'}=\pm i(\frac{1}{2}+p_L), \lambda_{p_R}=\pm i(\frac{1}{2}+p_R)$ to the Bethe equations corresponding to all spin down Bethe reference state \ref{BaeAdown}. Adding $\lambda_{p_L}$ we have

\bea\nonumber
(2N+1)a_{\frac{1}{2}}(\lambda)-a_{\frac{1}{2}+p_L}(\lambda)-a_{\frac{1}{2}+p_R}(\lambda)-a_{\frac{3}{2}+p_L}(\lambda)-a_{\frac{1}{2}-p_L}(\lambda)=\pi\delta(\lambda)+2\pi\rho_{|0\rangle_{p_L'}}(\lambda)+2\int\rho_{|0\rangle_{p_L'}}(\mu)a_1(\lambda-\mu)d\mu \\ \label{intAdownpL}\eea

Taking Fourier transform we obtain,

\bea
\label{denAdownpL}
\tilde{\rho}_{|0\rangle_{p_L'}}(\omega)=\tilde{\rho}_{|-\frac{1}{2}\rangle}(\omega)+\Delta \tilde{\rho}_{p_L'}(\omega), \;\;\; \Delta \tilde{\rho}_{p_L'}(\omega)=-\frac{e^{(\frac{3}{2}+p_L)|\omega |}+e^{(\frac{1}{2}-p_L)|\omega |}}{2(1+e^{-|\omega |})}
\eea

The number of real roots is given by $M_{|0\rangle{p_L'}}-1=\tilde{\rho}_{|0\rangle_{p_L'}}(0)$. From this we obtain the total number of roots $M_{|0\rangle{p_L'}}=\frac{N}{2}$, hence $S^z=0$. We observe that the number of roots is an integer only if the number of sites is even. Since we have a chain with odd number of sites, in order for one to add a boundary string to the state $|-\frac{1}{2}\rangle$, a propagating hole (spinon) needs to be added as well. Adding a spinon with rapidity $\theta$, to the state $|-\frac{1}{2}\rangle$ in addition to the boundary string $\lambda_{p_L'}$ we have

 \bea
 \label{intAdownpLhole}
 (2N+1)a_{\frac{1}{2}}(\lambda)-a_{\frac{1}{2}+p_L}(\lambda)-a_{\frac{1}{2}+p_R}(\lambda)-a_{\frac{3}{2}+p_L}(\lambda)-a_{\frac{1}{2}-p_L}(\lambda)\\=\pi\delta(\lambda)+\pi\delta(\lambda-\theta)+\pi\delta(\lambda+\theta)+2\pi\rho_{|-\frac{1}{2}\rangle_{(\theta,p_L')}}(\lambda)+2\int\rho_{|-\frac{1}{2}\rangle_{(\theta,p_L')}}(\mu)a_1(\lambda-\mu)d\mu
 \eea

Taking Fourier transform we obtain

\bea
\label{denAdownpLhole}
\tilde{\rho}_{|-\frac{1}{2}\rangle_{(\theta,p_L')}}(\omega)=\tilde{\rho}_{|-\frac{1}{2}\rangle}(\omega)+\Delta \tilde{\rho}_{p_L'}(\omega)+\Delta \tilde{\rho}_{\theta}(\omega), \;\;\;\; \Delta \tilde{\rho}_{\theta}(\omega)=-\frac{\cos(\theta\omega)}{(1+e^{-|\omega|})}\eea

The number of real roots is given by $M_{|0\rangle{(\theta,p_L')}}-1=\tilde{\rho}_{|0\rangle_{(\theta,p_L')}}(0)$. From this we find that the total number of Bethe roots is $M_{|-\frac{1}{2}\rangle_{(\theta,p_L')}}=\frac{N-1}{2}$, hence $S^z=-\frac{1}{2}$. We represent this state by $|-\frac{1}{2}\rangle_{(\theta,L)}$. We can calculate the energy of this state using \ref{energyAdown}, we have

 \bea
E_{|-\frac{1}{2}\rangle_{(\theta,p_L')}}= -4\int a_{\frac{1}{2}}(\lambda)\rho_{|-\frac{1}{2}\rangle_{(\theta,L)}}(\lambda)+N-1-\frac{1}{p_L}-\frac{1}{p_R}-\frac{2}{\frac{1}{4}+(i(\frac{1}{2}+p_L))^2}
\eea

Using \ref{denAdownpLhole}, in the above equation we obtain

\bea 
\label{energyAdownpLhole}
E_{|-\frac{1}{2}\rangle_{(\theta,L)}}=E_0+\frac{2\pi}{\sin(\pi p_L)}+\frac{2\pi}{\cosh(\theta)}.\eea

The first term is just the energy of the ground state $|-\frac{1}{2}\rangle$. The second term is the energy of the bound state at the left edge and the third term is the energy of the spinon with rapidity $\theta$. The energy of both the spinon and the bound state are strictly positive with 

$$0<E_{\theta}<2\pi ,\;\;\;\; E_{\text{bound state}}>2\pi$$

Similarly we can add the boundary string corresponding to the right boundary $\lambda_{p_R'}$ along with a spinon and obtain the state $|-\frac{1}{2}\rangle_{(\theta,R)}$ with total spin $S^z=-\frac{1}{2}$ described by the following density distribution

\bea
\label{denAdownpLhole}
\tilde{\rho}_{|-\frac{1}{2}\rangle_{(\theta,p_R')}}(\omega)=\tilde{\rho}_{|-\frac{1}{2}\rangle}(\omega)+\Delta \tilde{\rho}_{p_R'}(\omega)+\Delta \tilde{\rho}_{\theta}(\omega), \;\;\;\; \Delta \tilde{\rho}_{\theta}(\omega)=-\frac{\cos(\theta\omega)}{(1+e^{-|\omega|})}\eea

with energy

\bea 
\label{energyAdownpLhole}
E_{|-\frac{1}{2}\rangle_{(\theta,R)}}=E_0+\frac{2\pi}{\sin(\pi p_R)}+\frac{2\pi}{\cosh(\theta)}.\eea

\vspace{5mm}

Now consider the Bethe equations corresponding to all spin up reference state which can be obtained by taking $p_L\rightarrow -p_L$, $p_R\rightarrow -p_R$ \cite{Sklyannin}  in the equations \ref{BaeAdown}, \ref{BaeEAdown}.

\bea
\label{BaeAup}
\left(\frac{\lambda_j-i/2}{\lambda_j+i/2}\right)^{2N}\left(\frac{\lambda_j+i(\frac{1}{2}-p_{L})}{\lambda_j-i(\frac{1}{2}-p_{L})}\right)\left(\frac{\lambda_j+i(\frac{1}{2}-p_{R})}{\lambda_j-i(\frac{1}{2}-p_{R})}\right)\nonumber=\prod_{j\neq k=1}^{M}\left(\frac{\lambda_j-\lambda_k-i}{\lambda_j-\lambda_k+i}\right)\left(\frac{\lambda_j+\lambda_k-i}{\lambda_j+\lambda_k+i}\right)\eea

  The Eigenvalues of the Hamiltonian are given by 
 \bea 
\label{BaeEAup}
 E=-\sum_{j=1}^{M}\frac{2}{\lambda_j^2+\frac{1}{4}}+N-1+\frac{1}{p_L}+\frac{1}{p_R}
 \eea

By applying logarithm and following the same procedure as above, we obtain the following distribution for the state with all real Bethe roots.

\bea
 \label{denAupreal}
 \tilde{\rho}_{|\frac{1}{2}\rangle}(\omega)=\frac{(2N+1)e^{-\frac{|\omega|}{2}}-e^{-(\frac{1}{2}-p_L)|\omega|}-e^{-(\frac{1}{2}-p_R)|\omega|}-1}{2\left(1+e^{-|\omega|}\right)}
 \eea

The total number of roots is given by $M_{|\frac{1}{2}\rangle}=\tilde{\rho}_{|\frac{1}{2}\rangle}(0)$. Using which we obtain $M_{|\frac{1}{2}\rangle}=\frac{N-1}{2}$. Using this we obtain $S^z=\frac{1}{2}$.  We represent this state by $|\frac{1}{2}\rangle$. By using \ref{BaeAup} and \ref{denAupreal} we obtain the following expression for the energy of the state $|\frac{1}{2}\rangle$

\bea 
E_{|\frac{1}{2}\rangle}=\frac{2\pi}{\sin(\pi p_L)}+\frac{2\pi}{\sin (\pi p_R)}+E_{|-\frac{1}{2}\rangle}
\eea

Hence we find that this state contains bound states at the left and right edges, hence we represent this state by $|\frac{1}{2}\rangle_{L,R}$.

There exists states $|\frac{1}{2}\rangle_{(\theta,L)}$, $|\frac{1}{2}\rangle_{(\theta,R)}$ that are degenerate (in thermodynamic limit) to the states $|-\frac{1}{2}\rangle_{(\theta,L)}$, $|-\frac{1}{2}\rangle_{(\theta,R)}$ respectively obtained above. 

The state $|\frac{1}{2}\rangle_{(\theta,L)}$ contains a bound state at the left edge and a spinon. This state is obtained by adding the boundary string $\lambda_{p_R}=\pm i(\frac{1}{2}-p_R)$ which is a solution to \ref{BaeAup} and a spinon with rapidity $\theta$. Following the same procedure as above we obtain the following density distribution

\bea
\label{denAuppRhole}
\tilde{\rho}_{|\frac{1}{2}\rangle_{(\theta,p_R)}}(\omega)=\tilde{\rho}_{|\frac{1}{2}\rangle}(\omega)+\Delta \tilde{\rho}_{p_R}(\omega)+\Delta \tilde{\rho}_{\theta}(\omega), \;\;\;\; \Delta \tilde{\rho}_{p_R}(\omega)=-\frac{e^{(\frac{3}{2}-p_R)|\omega |}+e^{(\frac{1}{2}+p_R)|\omega |}}{2(1+e^{-|\omega |})}
\eea

The number of roots is given by $M_{|\frac{1}{2}\rangle_{(\theta,p_R)}}=\tilde{\rho}_{|\frac{1}{2}\rangle_{(\theta,p_R)}}(0)$. Using which we obtain $M_{|\frac{1}{2}\rangle_{(\theta,p_R}}=\frac{N-1}{2}$. Using this we obtain $S^z=\frac{1}{2}$.  The energy of this state can be obtained following the same procedure as above, we obtain

\bea 
\label{energyAdownpLhole}
E_{|\frac{1}{2}\rangle_{(\theta,L)}}=E_{|\frac{1}{2}\rangle}-\frac{2\pi}{\sin(\pi p_R)}+\frac{2\pi}{\cosh(\theta)}\equiv E_{-|\frac{1}{2}\rangle}+\frac{2\pi}{\sin(\pi p_L)}+\frac{2\pi}{\cosh(\theta)}.\eea

Hence it contains a bound state at the left edge and is degenerate with the state $|-\frac{1}{2}\rangle_{(\theta,L)}$ obtained previously. We represent this state by $|\frac{1}{2}\rangle_{(\theta,L)}$. Similarly, by adding the boundary string $\lambda_{p_L}=\pm i(\frac{1}{2}-p_L)$ and a spinon to the state $|\frac{1}{2}\rangle$, we obtain the state $|\frac{1}{2}\rangle_{(\theta,R)}$.

 Note that the bound state and the spinon both carry spin $\frac{1}{2}$. When a bound state at either the left or the right edge is added to the state which has spin $S^z=-\frac{1}{2}$, the bound state's spin is oriented in the positive $z$ direction. Now when a spinon is added it's spin can be oriented along or opposite to that of the bound state, and hence the final resulting state $|\pm\frac{1}{2}\rangle_{(\theta,j)}$, where $j=L,R$, has total spin $S^z=\pm\frac{1}{2}$ depending on the spin orientation of the spinon.
 
 \vspace{2mm}
 
 From the above analysis we see that for the odd chain, $|-\frac{1}{2}\rangle$ is the ground state. One can add one bound state at either the left edge or the right edge accompanied by a spinon and one obtains the states  $|\pm\frac{1}{2}\rangle_{(\theta,j)}$, where $j=L,R$. On can also add two bound states one at each edge and one obtains the state  $|\frac{1}{2}\rangle_{L,R}$.  Starting from either of these six states, one can build up excitations in the bulk by adding even number of spinons and other type of Bethe roots such as 2-strings and quartets \cite{DestriLowenstein}.

Here we summarize the construction of the ground state and the excited states in the phase $A_1$ for odd number of sites. In the region $\it{A}_1$ magnetic fields at both the boundaries point in the positive $z$ direction. The ground state has total spin $S^z=-\frac{1}{2}$ with energy $E_0$, exact expression of which is given by \ref{energygroundAdown} and is represented by $|-\frac{1}{2}\rangle$. It contains $\frac{N-1}{2}$ all real Bethe roots and is constructed by starting with all spin down reference state. There exists a state $|\frac{1}{2}\rangle_{L,R}$ with total spin $S^z=\frac{1}{2}$ which contains one exponentially localized spin $S^z=\frac{1}{2}$ boundary bound state at each edge. It contains  $\frac{N-1}{2}$ all real Bethe roots and is constructed by starting with reference state with all spin up. This state has energy $E_0+m_R+m_L$, where $m_i=\frac{2\pi}{\sin(\pi p_i)}$ is the energy of the bound state. 

\vspace{1mm}

There exists a state with a bound state only at the right edge, represented by $|-\frac{1}{2}\rangle_{(\theta,R)}$. It has total spin $S^z=-\frac{1}{2}$ and is obtained from the state $|-\frac{1}{2}\rangle$ by adding imaginary Bethe root $\lambda_{p_{R'}}=\pm i(\frac{1}{2}+p_R)$, which is called a boundary string. One also needs to add a spinon with rapidity $\theta$ in order to include the boundary string. There exists another state with a bound state at the right edge represented by $|\frac{1}{2}\rangle_{(\theta,R)}$. This state has total spin $S^z=\frac{1}{2}$, and is obtained from the state $|\frac{1}{2}\rangle_{L,R}$ by adding the boundary string $\lambda_{p_{L}}=\pm i(\frac{1}{2}-p_L)$ and also a spinon with rapidity $\theta$. The two states $|\frac{1}{2}\rangle_{(\theta,R)},|-\frac{1}{2}\rangle_{(\theta,R)}$ are degenerate in the thermodynamic limit and have energy $E_0+E_{\theta}+E_{R}$ but differ in the spin orientation of the spinon which has spin $S^z=\pm \frac{1}{2}$ respectively. Here $E_{\theta}=\frac{2\pi}{\cosh(\pi \theta)}$ is the energy of the spinon with rapidity $\theta$.

\vspace{1mm}

There also exists two degenerate states $|\pm\frac{1}{2}\rangle_{(\theta,L)}$, with energy $E_0+m_L+E_{\theta}$.  The state $|-\frac{1}{2}\rangle_{(\theta,L)}$ has total spin $S^z=-\frac{1}{2}$ and can be obtained from the state $|-\frac{1}{2}\rangle$ by adding the boundary string $\lambda_{p_{L'}}=\pm i(\frac{1}{2}+p_L)$ and a spinon with rapidity $\theta$, whose spin is $S^z=-\frac{1}{2}$. The state $|\frac{1}{2}\rangle_{(\theta,L)}$ has total spin $S^z=\frac{1}{2}$, and is obtained from the state $|\frac{1}{2}\rangle$ by adding the boundary string $\lambda_{p_{R}}=\pm i(\frac{1}{2}-p_R)$ and a spinon with rapidity $\theta$ whose spin is $S^z=\frac{1}{2}$.

\subsubsection{$A_1$: Even number of sites}

Now consider the spin chain with even number of sites. As seen in the previous section, starting with all spin down reference state and considering a state with all real roots one obtains the ground state $|-\frac{1}{2}\rangle$. The number of roots in this state is $M=\frac{N-1}{2}$. N has to be odd in order for the number of roots to be an integer. To obtain the ground state for even number of sites, one needs to consider a state with one less Bethe root compared to the state $|-\frac{1}{2}\rangle$, that is starting with all spin down reference state we need to include one spinon in addition to all real Bethe roots. Following the procedure described in the previous section, we obtain the following distribution

\bea
\label{denAdownholeven}
\tilde{\rho}_{|-1\rangle_{\theta}}(\omega)=\frac{(2N+1)e^{-\frac{|\omega|}{2}}-e^{-(\frac{1}{2}+p_L)|\omega|}-e^{-(\frac{1}{2}+p_R)|\omega|}-1}{2\left(1+e^{-|\omega|}\right)}
+\Delta \tilde{\rho}_{\theta}(\omega), \;\;\;\; \Delta \tilde{\rho}_{\theta}(\omega)=-\frac{\cos(\theta\omega)}{(1+e^{-|\omega|})}\eea

The total number of Bethe roots is $M_{|-1\rangle_{\theta}}=\frac{N-2}{2}$, hence $S^z=-1$. The number of roots is an integer only for a spin chain with even number of sites as desired. Note that the first term is same as the density distribution describing the state $|-\frac{1}{2}\rangle$, with $N$ now being even. The energy of this state can be calculated using \ref{BaeEAdown}, we obtain

\bea E_{|1\rangle_{\theta}}=E_0+ \frac{2\pi}{\cosh(\theta)}.\eea

Hence in $A_1$, the lowest energy state for even number of sites chain is parametrized by the rapidity of the  spinon $\theta$. The ground state is obtained in the limit where $\theta\rightarrow \infty$. Starting with all spin up reference state and considering a state with all real roots and a spinon we obtain

\bea
\label{denAupholeven}
\tilde{\rho}_{|1\rangle_{\theta}}(\omega)=\frac{(2N+1)e^{-\frac{|\omega|}{2}}-e^{-(\frac{1}{2}-p_L)|\omega|}-e^{-(\frac{1}{2}-p_R)|\omega|}-1}{2\left(1+e^{-|\omega|}\right)}
+\Delta \tilde{\rho}_{\theta}(\omega)\eea

The total number of Bethe roots is $M_{|1\rangle_{\theta}}=\frac{N-2}{2}$, hence $S^z=1$. The energy of this state can be calculated using \ref{BaeEAup}, we obtain

\bea 
E_{|1\rangle_{\theta}}=E_0+\frac{2\pi}{\sin(\pi p_L)}+\frac{2\pi}{\sin (\pi p_R)}+\frac{2\pi}{\cosh(\theta)}.\eea

This state contains bound states at both the edges, and hence we represent this state by $|1\rangle_{(\theta,L,R)}$. Starting with all spin down reference state, consider a state with all real roots and the boundary string $\lambda_{p_L'}=\pm i(\frac{1}{2}+p_L)$. Following the similar procedure described in the previous section we obtain

\bea
\label{denAdownpLeven}
\tilde{\rho}_{|0\rangle_{p_L'}}(\omega)=\frac{(2N+1)e^{-\frac{|\omega|}{2}}-e^{-(\frac{1}{2}+p_L)|\omega|}-e^{-(\frac{1}{2}+p_R)|\omega|}-1}{2\left(1+e^{-|\omega|}\right)}+\Delta \tilde{\rho}_{p_L'}(\omega), \;\;\; \Delta \tilde{\rho}_{p_L'}(\omega)=-\frac{e^{(\frac{3}{2}+p_L)|\omega |}+e^{(\frac{1}{2}-p_L)|\omega |}}{2(1+e^{-|\omega |})}
\eea

The total number of roots is $M_{|0\rangle_{p_L'}}=\frac{N}{2}$, and hence $S^z=0$. We represent this state by $|0\rangle_{L}$. The energy of this state can be calculated using the procedure described in the previous section. We obtain

\bea  E_{|0\rangle_{L}}=E_0+\frac{2\pi}{\sin(\pi p_L)} \eea

Similarly we can add the boundary string $\lambda_{p_R'}=\pm i(\frac{1}{2}+p_R)$ and obtain the state $|0\rangle_R$ which has energy $ E_{|0\rangle_{R}}=E_0+\frac{2\pi}{\sin(\pi p_R)} $. These states can also be obtained by starting with all spin up reference state, in which case, $|0\rangle_{R}$, $|0\rangle_L$ contain all real roots and the boundary strings $\lambda_{p_L}=\pm i(\frac{1}{2}-p_L)$, $\lambda_{p_R}=\pm i(\frac{1}{2}-p_R)$ respectively.

Starting with all spin down reference state, one can add both the boundary strings $\lambda_{p_R'}, \lambda_{p_L'}$ to the state with all real roots and one spinon. Following the regular procedure, we obtain the following density distribution

\bea
\label{denAdownpLpReven}
\tilde{\rho}_{|0\rangle_{(\theta p_L',p_R')}}(\omega)=\frac{(2N+1)e^{-\frac{|\omega|}{2}}-e^{-(\frac{1}{2}+p_L)|\omega|}-e^{-(\frac{1}{2}+p_R)|\omega|}-1}{2\left(1+e^{-|\omega|}\right)}+\Delta \tilde{\rho}_{p_L'}(\omega) + \Delta \tilde{\rho}_{p_R'}(\omega)+ \Delta \tilde{\rho}_{\theta}(\omega)\eea

The number of real roots is given by $M_{|0\rangle_{(\theta p_L',p_R')}}-2=\tilde{\rho}_{|0\rangle_{(\theta p_L',p_R')}}(0)$. From this we obtain that the total number of roots is $M_{|0\rangle_{(\theta p_L',p_R')}}=\frac{N}{2}$. This results in $S^z=0$. We represent this state by $|0\rangle_{(\theta,L,R)}$. Calculating the energy we find that this state is degenerate with the state $|1\rangle_{(\theta, L,R)}$. Similarly, we can add both the boundary strings $\lambda_{p_R}, \lambda_{p_L}$ to the state $|1\rangle_{\theta}$ and obtain the state $0\rangle_{\theta}$ which is degenerate with the state $|-1\rangle_{\theta}$.

Here we summarize the construction of the ground state and the excited states in the phase $A_1$ for even number of sites. In the phase $A_1$, the ground state $|-1\rangle_{\theta}$ has total spin $S^z=-1$. It is constructed by starting with all spin down reference state and contains $\frac{N-2}{2}$ real roots and a spinon with rapidity $\theta$. In the thermodynamic limit there exists another state $|0\rangle_{\theta}$ with total spin $S^z=0$ which is degenerate with the ground state. This state is obtained by starting with all spin up reference state and in addition to $\frac{N-4}{2}$ real roots it contains two boundary strings $\lambda_{p_{R}}$, $\lambda_{p_{L}}$ and a spinon with rapidity $\theta$. The spin orientation of the spinon in the two ground states $|-1\rangle_{\theta}$, $|0\rangle_{\theta}$ is along the negative and positive $z$ directions respectively.

\vspace{1mm}

There exists a state represented by $|0\rangle_L$ which contains a bound state at the left edge and has total spin $S^z=0$. This is constructed from the reference state with all spin down. In addition to $\frac{N-2}{2}$ real roots it contains one boundary string $\lambda_{p_{L'}}$ and has energy $E_0+m_L$. This state can also be constructed from the reference state with all spin up and is made up of $\frac{N-2}{2}$ real roots and one boundary string $\lambda_{p_R}$. There exists a state $|0\rangle_{R}$ which contains a bound state at the right edge and has total spin $S^z=0$. It is constructed from the reference state with all spin down and in addition to $\frac{N-2}{2}$real roots it contains one boundary string $\lambda_{p_{R'}}$ and has energy $E_0+m_R$. This state can also be constructed from the reference state with all spin up and in addition to $\frac{N-2}{2}$ real roots it contains one boundary string $\lambda_{p_L}$.

\vspace{1mm}

There also exists a state $|0\rangle_{(\theta,L,R)}$ with total spin $S^z=0$ which contains one bound state at each edge. It is constructed by starting with all spin down reference state and contains $\frac{N-4}{2}$ real roots and two boundary strings $\lambda_{p_{L'}}$, $\lambda_{p_{R'}}$ and a spinon. The energy of the state is $E_0+m_R+m_L+E_{\theta}$. In the thermodynamic limit there exists another degenerate state $|1\rangle_{(\theta,L,R)}$ with total spin $S^z=1$ that contains one bound state at each edge. This state is constructed by starting with all spin up reference state and contains $\frac{N-2}{2}$ real roots and a spinon. The spin orientation of the spinon in the two states $|0\rangle_{(\theta,L,R)}$ and $|1\rangle_{(\theta,L,R)}$ is in the negative and positive $z$ directions respectively.

\subsubsection{Phase $A_2$ }
Construction of the states in the phase $A_2$ is similar to that in the $A_1$ phase, hence we skip the details of the calculation and provide an overview of how the ground state and boundary excited states are constructed in $A_2$ phase. The Bethe equations for all down and up reference states are obtained from \ref{BaeAdown}, \ref{BaeEAdown} and \ref{BaeAup}, \ref{BaeEAup} by making the transformation $p_L\rightarrow -p_L$. 

\vspace{2mm}

Consider the case of odd number of sites. The ground state is two fold degenerate with energy $E_0+E_{\theta}$ and is represented by $|\pm\frac{1}{2}\rangle_{\theta}$ with total spin $S^z=\pm\frac{1}{2}$ respectively.  $|-\frac{1}{2}\rangle_{\theta}$ is constructed from all spin down reference state. It contains the boundary string $\lambda_{p_L}$ and a spinon in addition to $\frac{N-3}{2}$ real roots. The state $|\frac{1}{2}\rangle_{\theta}$ is constructed by starting with all spin up reference state. It contains the boundary string $\lambda_{p_R}$ and a spinon in addition to $\frac{N-3}{2}$ real roots.  The state with a bound state at the left edge is represented by $|-\frac{1}{2}\rangle_L$ and has total spin $S^z=-\frac{1}{2}$ and energy $E_0+m_L$. It is constructed by starting with all spin down reference state and contains $\frac{N-1}{2}$ all real roots. The state with a bound state at the right edge is represented by $|\frac{1}{2}\rangle_R$ and has total spin $S^z=\frac{1}{2}$ and energy $E_0+m_R$. It is constructed by starting with all spin up reference state and contains $\frac{N-1}{2}$ all real roots. The state with bound states at both the edges is two fold degenerate with energy $E_0+m_L+m_R+E_{\theta}$ and is represented by $|\pm\frac{1}{2}\rangle_{(\theta,L,R)}$ with total spin $S^z=\pm\frac{1}{2}$ respectively. $|\frac{1}{2}\rangle_{(\theta,L,R)}$ is constructed by starting with all spin up reference state and contains the boundary string $\lambda_{p_L'}$ and a  spinon in addition to $\frac{N-3}{2}$ real roots. $|-\frac{1}{2}\rangle_{(\theta,L,R)}$ is constructed by starting with all spin down reference state and contains the boundary string $\lambda_{p_R'}$ and a spinon in addition to $\frac{N-3}{2}$ real roots.

\vspace{2mm}

Now consider the case of even number of sites. The ground state is represented by $|0\rangle$ and has energy $E_0$ and total spin $S^z=0$. It is constructed by starting with all spin up reference state and contains the boundary string $\lambda_{p_R}$ in addition to $\frac{N-2}{2}$ real roots. It can also be constructed by starting with all spin down reference state and contains the boundary string $\lambda_{p_L'}$ in addition to $\frac{N-2}{2}$ real roots. The state which contains a bound state at the left edge is two fold degenerate with energy $E_0+m_L+E_{\theta}$ and is represented by $|0\rangle_{(\theta,L)}$ and $|-1\rangle_{(\theta,L)}$ with total spin $S^z=0$  and $S^z=-1$ respectively. The state $|0\rangle_{(\theta,L)}$ is constructed by starting with all spin up reference state. It contains the boundary strings $\lambda_{p_R}$, $\lambda_{p_L'}$ and a spinon in addition to $\frac{N-4}{2}$ real roots. The state $|-1\rangle_{(\theta,L)}$ is constructed by starting with all spin down reference state and contains a spinon and $\frac{N-2}{2}$ real roots. The state which contains a bound state at the right edge is two fold degenerate with energy $E_0+m_R+E_{\theta}$ and is represented by $|0\rangle_{(\theta,R)}$ and $|1\rangle_{(\theta,R)}$ with total spin $S^z=0$  and $S^z=1$ respectively. The state $|0\rangle_{(\theta,R)}$ is constructed by starting with all spin down reference state. It contains the boundary strings $\lambda_{p_R'}$, $\lambda_{p_L}$ and a spinon in addition to $\frac{N-4}{2}$  real roots. The state $|1\rangle_{(\theta,R)}$ is constructed by starting with all spin up reference state and contains a spinon and $\frac{N-2}{2}$ real roots. The state with bound states at both the edges is represented by $|0\rangle_{L,R}$ and has total spin $S^z=0$ and energy $E_0+m_L+m_R$. It is constructed by starting with all spin up reference state and contains the boundary string $\lambda_{p_L'}$ in addition to $\frac{N-2}{2}$ real roots. It can also be constructed by starting with all spin down reference state and contains the boundary string $\lambda_{p_R'}$ in addition to $\frac{N-2}{2}$ real roots.

\subsubsection{Phase $A_3$, $A_4$}
 In the phases $A_3$, $A_4$, both the boundary magnetic fields point in the direction opposite that of the phases $A_1$ and $A_2$ respectively. Using the property \ref{z2}, we can obtain all the states in the phases $A_3$ and $A_4$ from the states obtained in the phase $A_1$ and $A_2$ respectively. 
\vspace{2mm}

In constructing a state in the phase $A_3$ or $A_4$, we can use the construction of the respective state in the phase $A_1$ or $A_2$ respectively, and use the following transformation:

\bea\label{z2bethe}
 |\uparrow\uparrow...\uparrow\rangle \leftrightarrow |\downarrow\downarrow...\downarrow\rangle, \hspace{5mm} p_L\rightarrow -p_L, \hspace{4mm}p_R\rightarrow p_R \eea

where the all spin up and all spin down reference states are interchanged and the boundary magnetic fields change sign.

\subsection{$B$ phases}
Consider the phase $B_1$. In this phase both the magnetic fields point in the positive $z$- direction and take the values $0<h_R<h_C$ and $h_L>h_C$, which corresponds to the values $p_R>\frac{1}{2}$ and $0<p_L<\frac{1}{2}$.

\subsubsection{Phase $B_1$}

Consider the chain with odd number of sites. The Bethe equations corresponding to all spin down reference state are given by \ref{BaeAdown}. The Eigenvalues of the Hamiltonian are given by \ref{BaeEAdown}. Ground state contains all real roots and we obtain the distribution \ref{denAdownreal} with energy $E_0$ and spin $S^z=-\frac{1}{2}$. Now by adding the boundary string $\lambda_{p_L'}$ along with a spinon we obtain the state $|-\frac{1}{2}\rangle_{(\theta,L)}$ which contains a bound state at the left edge. This state has total spin $S^z=-\frac{1}{2}$ and energy $E_0+m_L+E_{\theta}$.

Now consider all spin up reference state. The eigenvalues are given by \ref{BaeEAup} where as the Bethe equations \ref{BaeAup} take the form

\bea
\label{BaeBup}
\left(\frac{\lambda_j-i/2}{\lambda_j+i/2}\right)^{2N}\left(\frac{\lambda_j+i(\frac{1}{2}-p_{L})}{\lambda_j-i(\frac{1}{2}-p_{L})}\right)\left(\frac{\lambda_j-i(p_R-\frac{1}{2})}{\lambda_j+i(p_R-\frac{1}{2})}\right)\nonumber=\prod_{j\neq k=1}^{M}\left(\frac{\lambda_j-\lambda_k-i}{\lambda_j-\lambda_k+i}\right)\left(\frac{\lambda_j+\lambda_k-i}{\lambda_j+\lambda_k+i}\right)\eea

Considering the state with all real roots and following the same procedure as in the previous section we obtain

\bea
 \label{denBupreal}
 \tilde{\rho}_{|0\rangle}(\omega)=\frac{(2N+1)e^{-\frac{|\omega|}{2}}-e^{-(\frac{1}{2}-p_L)|\omega|}+e^{-(p_R-\frac{1}{2})|\omega|}-1}{2\left(1+e^{-|\omega|}\right)}
 \eea

The total number of roots given by $M=\tilde{\rho}_{|0\rangle}(0)=\frac{N}{2}$. Hence we find that the number of roots is not an integer for odd number of sites. Hence we need to consider the state with one spinon along with real roots. We obtain

\bea
 \label{denBuprealhole}
 \tilde{\rho}_{|\frac{1}{2}\rangle_{\theta}}(\omega)=\frac{(2N+1)e^{-\frac{|\omega|}{2}}-e^{-(\frac{1}{2}-p_L)|\omega|}+e^{-(p_R-\frac{1}{2})|\omega|}-1}{2\left(1+e^{-|\omega|}\right)}+\Delta \tilde{\rho}_{\theta}(\omega) \eea

The number of roots is given by $M= \tilde{\rho}_{|\frac{1}{2}\rangle_{\theta}}(0)=\frac{N-1}{2}$, and hence the number of roots is an integer for odd number of sites and we obtain $S^z=\frac{1}{2}$. The energy of this state can be calculated using \ref{BaeEAup}. Following the procedure described in the previous section we obtain $ E_{|\frac{1}{2}\rangle_{\theta}}=E_0+m_L+E_{\theta}$. Hence this state contains a bound state at the left edge and is degenerate with the state $|-\frac{1}{2}\rangle_{(\theta,L)}$ obtained above. We represent this state by $|\frac{1}{2}\rangle_{(\theta,L)}$. 

\vspace{2mm}

Now consider the chain with even number of sites. The Bethe equation corresponding to all spin up reference state have the boundary string solution $\lambda_{p_L}$. Consider a state with this boundary string and a spinon in addition to real roots. By following the usual procedure we obtain the following distribution

\bea
 \label{denBuprealholepL}
 \tilde{\rho}_{|0\rangle_{(\theta,p_L)}}(\omega)=\frac{(2N+1)e^{-\frac{|\omega|}{2}}-e^{-(\frac{1}{2}-p_L)|\omega|}+e^{-(p_R-\frac{1}{2})|\omega|}-1}{2\left(1+e^{-|\omega|}\right)}+ \Delta \tilde{\rho}_{p_L}(\omega)+\Delta \tilde{\rho}_{\theta}(\omega) \eea

The number of real roots is given by $M_{|0\rangle_{(\theta,p_L)}}-1= \tilde{\rho}_{|0\rangle_{(\theta,p_L)}}(0)=\frac{N-2}{2}$, hence we obtain that the total number of roots is $M_{|0\rangle_{(\theta,p_L)}}=\frac{N}{2}$ and hence $S^z=0$. The energy can be calculated using \ref{BaeEAup}, we obtain $E_{|0\rangle_{(\theta,p_L)}}=E_0+E_{\theta}$. Starting with the spin down reference state, consider the state with one spinon in addition to real roots. We obtain the following distribution

\bea
\label{denBdownrealhole}
\tilde{\rho}_{|-1\rangle_{\theta}}(\omega)=\frac{(2N+1)e^{-\frac{|\omega|}{2}}-e^{-(\frac{1}{2}-p_L)|\omega|}-e^{-(\frac{1}{2}-p_R)|\omega|}-1}{2\left(1+e^{-|\omega|}\right)}+\Delta \tilde{\rho}_{\theta}(\omega)\eea

The number of roots is given by $M_{|-1\rangle_{\theta}}=\frac{N-2}{2}$, hence $S^z=-1$. The energy of this state can be calculated using \ref{BaeEAdown}, and we obtain $E_{|-1\rangle_{\theta}}=E_0+E_{\theta}$. Hence the ground state is two fold degenerate with energy $E_0+E_{\theta}$. It is represented by $|0\rangle_{\theta}$ and $|-1\rangle_{\theta}$ with total spin $S^z=0$ and $S^z=-1$ respectively. 

\vspace{2mm}
 
 Starting with all spin up reference state, consider the state with all real roots. We obtain the following distribution
 
  \bea
 \label{denB1upreal}
 \tilde{\rho}_{|0\rangle}(\omega)=\frac{(2N+1)e^{-\frac{|\omega |}{2}}-e^{-(\frac{1}{2}-p_L)|\omega |}+e^{-(p_R-\frac{1}{2})|\omega |}-1}{2\left(1+e^{-|\omega |}\right)}
 \eea

The number of roots is given by $M_{|0\rangle}= \tilde{\rho}_{|0\rangle}(0)=\frac{N}{2}$, and hence it has total spin $S^z=0$. The energy of this state can be calculated using \ref{BaeEAup}, we obtain  $E_{|0\rangle}=E_0+m_L$. Hence it has a bound state at the left edge. This state is represented by $|0\rangle_L$.

In summary, for the spin chain with odd number of sites, in the phases $B_1$, the ground state has total spin $S^z=-\frac{1}{2}$ and is represented by $|-\frac{1}{2}\rangle$. This state is constructed by starting with all spin down  reference state and contains $\frac{N-1}{2}$ real roots and has energy $E_0$. There exists a state which contains a bound state at the left edge, which is represented by $|-\frac{1}{2}\rangle_{(\theta,L)}$ and has total spin $S^z=-\frac{1}{2}$ and energy $E_0+E_{\theta}+m_L$. This state is obtained from the state $|-\frac{1}{2}\rangle$ by adding the boundary string $\lambda_{p_{L'}}$ and a spinon with rapidity $\theta$. There exists a degenerate state $|\frac{1}{2}\rangle_{(\theta,L)}$ which contains a bound state at the left edge and has total spin $S^z=\frac{1}{2}$. This state is obtained by starting with reference state with all spin up. It contains $\frac{N-1}{2}$ real roots and a spinon with rapidity $\theta$.

For the spin chain with even number of sites, in the phase $B_1$, the ground state is two fold degenerate. The ground state $|-1\rangle_{\theta}$, with total spin $S^z=-1$ is constructed by starting with all spin down reference state and contains $\frac{N-2}{2}$ real roots and a spinon with rapidity $\theta$. The ground state $|0\rangle_{\theta}$ with total spin $S^z=0$ is constructed by starting with all spin up reference state. It contains a spinon and the boundary string $\lambda_{p_{L}}$ in addition to $\frac{N-2}{2}$ real roots. The spin orientation of the spinon in the states $|-1\rangle_{\theta}$ and $|0\rangle_{\theta}$ is along the negative and positive $z$ directions respectively. There exists a state $|0\rangle_{L}$ with a bound state at the left boundary. It has total spin $S^z=0$ and is constructed by starting with all spin up reference state and contains $\frac{N}{2}$ real roots. It can also be constructed by starting with all spin down reference state and it includes the boundary string $\lambda_{p_{L'}}$ in addition to $\frac{N-2}{2}$ real roots.

\subsubsection{Phase $B_2$}

The construction of the state in the phase $B_2$ is similar to that in the phase $B_1$, hence we skip the details of the calculation and provide an overview of how the ground state and boundary excited states are constructed in the phase $B_2$.

\vspace{2mm}

Consider the chain with odd number of sites. The ground state is two fold degenerate with energy $E_0+E_{\theta}$. It is represented by $|\pm\frac{1}{2}\rangle_{\theta}$ with total spin $S^z=\pm \frac{1}{2}$. The state $|\frac{1}{2}\rangle_{\theta}$ is obtained by starting with reference state with all spin up and it contains a spinon in addition to $\frac{N-2}{2}$ real Bethe roots. The state $|-\frac{1}{2}\rangle_{\theta}$ is obtained by starting with all spin down reference state and contains the boundary string $\lambda_{p_L}$ and a spinon in addition to real roots. The state with the bound state at the left edge has energy $E_0+m_L$ and total spin $S^z=-\frac{1}{2}$. It is represented by $|-\frac{1}{2}\rangle_{L}$ and is constructed by starting with all spin down reference state and contains all real roots.

\vspace{2mm}

Consider the chain with even number of sites. The ground state $|0\rangle$ has energy $E_0$ and total spin $S^z=0$. It is obtained by starting with reference state with all spin up and it contains $\frac{N}{2}$ all real roots. It can also be obtained by starting with all spin down reference state and it contains the boundary string $\lambda_{p_L}$ in addition to $\frac{N-2}{2}$real roots. The state with the bound state at the left edge is doubly degenerate with energy $E_0+m_L+E_{\theta}$ and is represented by $|-1\rangle_{(\theta,L)}$ and $|0\rangle_{(\theta,L)}$ with total spin $S^z=-1$ and $S^z=0$ respectively. The state $|-1\rangle_{(\theta,L)}$ is obtained by starting with all spin down reference state and contains one spinon in addition to $\frac{N-2}{2}$ real roots. The state $|0\rangle_{(\theta,L)}$ is obtained by starting with all spin all reference state. It contains the boundary string $\lambda_{p_L'}$ and a spinon in addition to $\frac{N-2}{2}$ real roots.

\subsubsection{Other $B$ phases}
The states in the phases $B_8$ and $B_7$ can be obtained from the states in the phases $B_1$ and $B_2$ respectively by the transformation $p_L\leftrightarrow p_R$.
The states in the phases $B_5$,$B_6$, $B_3$ and $B_4$ can be obtained from the states in the phases $B_1$, $B_2$, $B7$ and $B_8$ respectively by the transformation \ref{z2bethe}.

\subsection{$C$ phases}
\subsubsection{Odd number of sites}
 In subregion $\it{C_1}$, the ground state is $|-\frac{1}{2}\rangle$ with total spin $S^z=-\frac{1}{2}$. This state is constructed from the reference state with all down spins and contains $\frac{N-1}{2}$ real roots. In $\it{C_3}$ the ground state is $|\frac{1}{2}\rangle$ with total spin $S^z=\frac{1}{2}$. This state is constructed from the reference state with all up spins and contains $\frac{N-1}{2}$ real roots. In subregions $\it{C_2}, \it{C_4}$ the ground state is two fold degenerate and contains a spinon with rapidity $\theta$. The spin orientation of the spinon dictates the total spin $S^z=\pm\frac{1}{2}$ of the state. They are represented by $|\pm\frac{1}{2}\rangle_{(\pm,\theta)}$, and contain $\frac{N-1}{2}$ real roots and a spinon with rapidity $\theta$, and constructed from either all spin up or down reference states and contain all real roots and a spinon with rapidity $\theta$.

 \subsubsection{Even number of sites}
 In subregion $\it{C_1}$, the ground state is two fold degenerate in thermodynamic limit and is represented by $|0\rangle_{\theta}$, $|-1\rangle_{\theta}$ with total spin $S^z=0$, $S^z=-1$ respectively. The state $|0\rangle_{\theta}$ is constructed from the reference state with all up spin and contains $\frac{N}{2}$ real roots and a spinon with rapidity $\theta$. the state $|-1\rangle_{\theta}$ is constructed with the reference state with all down spins and contains $\frac{N-2}{2}$ real roots and a spinon with rapidity $\theta$. The spin orientation of the spinon is in the negative and positive $z$ direction in the states $|-1\rangle_{\theta}$ and $|0\rangle_{\theta}$ respectively. In subregion $\it{C_3}$, the ground state is two fold degenerate and is represented by $|0\rangle_{\theta}$, $|1\rangle_{\theta}$ with total spin $S^z=0$, $S^z=1$ respectively. $|0\rangle_{\theta}$ contains a spinon with rapidity $\theta$, the spin orientation of which is in the negative $z$ direction. It is constructed from the reference state with all spin down and contains $\frac{N}{2}$ real roots and a spinon with rapidity $\theta$. $|1\rangle_{\theta}$ is constructed with the reference state with all spin up and contains $\frac{N-2}{2}$ real roots and a spinon with rapidity $\theta$ with spin oriented in the positive $z$ direction. In subregions $C_2,C_4$, the ground state has total spin $S^z=0$ and is represented by $|0\rangle$. It can be constructed from the reference state with all spin up or all spin down and contains $\frac{N}{2}$ real roots. 

\section{Bound state wavefunction}
\label{appendix2}
In this section we provide the bound state wavefunction corresponding to the boundary string in one particle sector (one flipped spin). Let us consider the sub-phase $A_1$. The Bethe equations corresponding to all spin down reference state in one particle sector are given by

\bea
\label{oneBEA1}
\left(\frac{\lambda-\frac{i}{2}}{\lambda+\frac{i}{2}}\right)^{2N}\left(\frac{\lambda+i(\frac{1}{2}+p_{L})}{\lambda-i(\frac{1}{2}+p_{L})}\right)\left(\frac{\lambda+i(\frac{1}{2}+p_{R})}{\lambda-i(\frac{1}{2}+p_{R})}\right)=1
\eea

The wavefunction is given by \cite{Alcaraz,skorik}

\bea\label{WFboundaryR}
f(x)=\left(\frac{\lambda+\frac{i}{2}}{\lambda-\frac{i}{2}}\right)^{N-x}\left(\frac{\lambda-i\left(\frac{1}{2}+p_R\right)}{p_R\left(\lambda-\frac{i}{2}\right)}\right)-\left(\frac{\lambda-\frac{i}{2}}{\lambda+\frac{i}{2}}\right)^{N-x}\left(\frac{\lambda+i\left(\frac{1}{2}+p_R\right)}{p_R\left(\lambda+\frac{i}{2}\right)}\right)
\eea

When $\lambda=\pm i(\frac{1}{2}+p_R)$, which is the boundary string associated with the right edge, we readily obtain the wavefunction for the boundstate localized at the right edge

\bea\label{WFR}
f_R(x)=\pm \left(\frac{1+2p_R}{p_R(1+p_R)}\right)\left(\frac{1+p_R}{p_R}\right)^{-(N-x)}
\eea

To obtain the bound state wavefunction associated with the left edge, we multiply the wavefunction \eqref{WFboundaryR} with a normalization constant

\bea\label{normalizationconstant}
\mathcal{A}=\left(\frac{p_R}{p_L}\right)\left(\frac{\lambda+i(\frac{1}{2}+p_{L})}{\lambda+i(\frac{1}{2}+p_{R})}\right)\left(\frac{\lambda-i(\frac{1}{2}+p_{L})}{\lambda-i(\frac{1}{2}+p_{R})}\right)
\eea
 and use the one particle Bethe equation \eqref{oneBEA1}. We obtain
 
 \bea\label{WFboundaryL}
 \mathcal{A}f(x)=\left(\frac{\lambda+\frac{i}{2}}{\lambda-\frac{i}{2}}\right)^{-x}\left(\frac{\lambda+i\left(\frac{1}{2}+p_L\right)}{p_L\left(\lambda-\frac{i}{2}\right)}\right)-\left(\frac{\lambda-\frac{i}{2}}{\lambda+\frac{i}{2}}\right)^{-x}\left(\frac{\lambda-i\left(\frac{1}{2}+p_L\right)}{p_L\left(\lambda+\frac{i}{2}\right)}\right)
 \eea
 
When $\lambda=\pm i(\frac{1}{2}+p_L)$, which is the boundary string associated with the left edge, we obtain the bound state wavefunction localized at the left edge

\bea\label{WFR}
f_L(x)=\pm \left(\frac{1+2p_L}{p_L^2}\right)\left(\frac{1+p_L}{p_L}\right)^{-x}
\eea

Hence, we find that the two boundary string solutions correspond to exponentially localized bound states $\sim e^{-\kappa_L x}$ and $\sim e^{-\kappa_R (N-x)}$, where 
\bea \label{boundstatesexponetial}\kappa_j=\log \left(\frac{h_j+1}{h_j}\right), \;\;\; j=L,R. \eea

Note that when the magnetic fields at the edges take equal values $h_L=h_R=h$ ($p_L=p_R=p$), the normalization constant $\mathcal{A}=1$. To obtain the bound state wavefunctions in this case, consider 

\bea\label{doubleWFboundary}
f(x)\pm \left(\frac{p}{1+p}\right)f(x).
\eea

In this limit we only have one boundary string solution $\lambda=\pm i(\frac{1}{2}+p)$, corresponding to the double pole of the Bethe equations \eqref{oneBEA1}. Using this in \eqref{doubleWFboundary}, we obtain 

\bea\label{doubleWF}
f(x)_{\pm}=-\left(\frac{1+2p}{p(1+p)}\right)\left[\left(\frac{1+p}{p}\right)^{-(N-x)}\pm \left(\frac{1+p}{p}\right)^{-x}\right]
\eea

Hence, we find that when $h_L=h_R$, in contrast to \cite{skorik}, there exist two bound states solutions simultaneously localized at both the edges 

\bea f(x)_{\pm}\sim (e^{-\kappa(N-x)}\pm e^{-\kappa x}), \;\;\; \kappa=\log \left(\frac{h+1}{h}\right). \eea


\section{Details and validity of the fitting}
\label{sec:fitting}

\begin{figure*}
    \centering
    \includegraphics[width=0.34\linewidth]{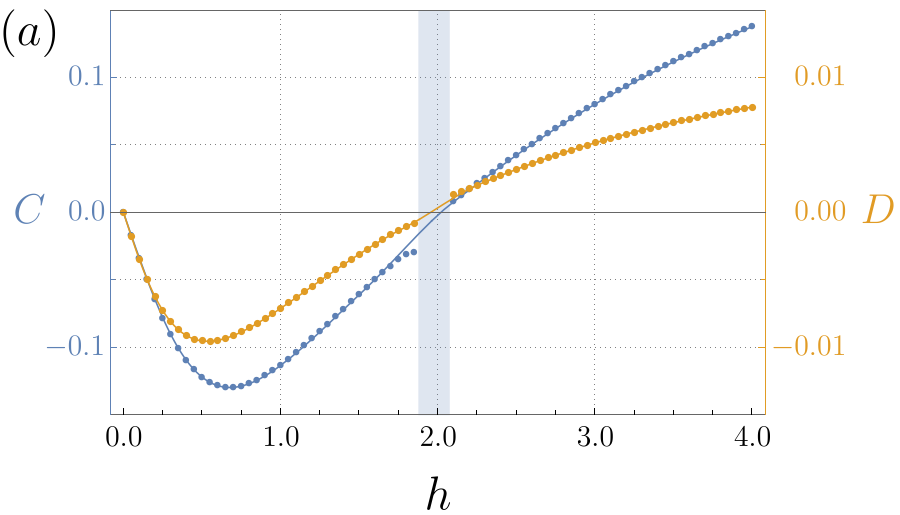}
    \includegraphics[width=0.34\linewidth]{fitParams2.png}
    \includegraphics[width=0.30\linewidth]{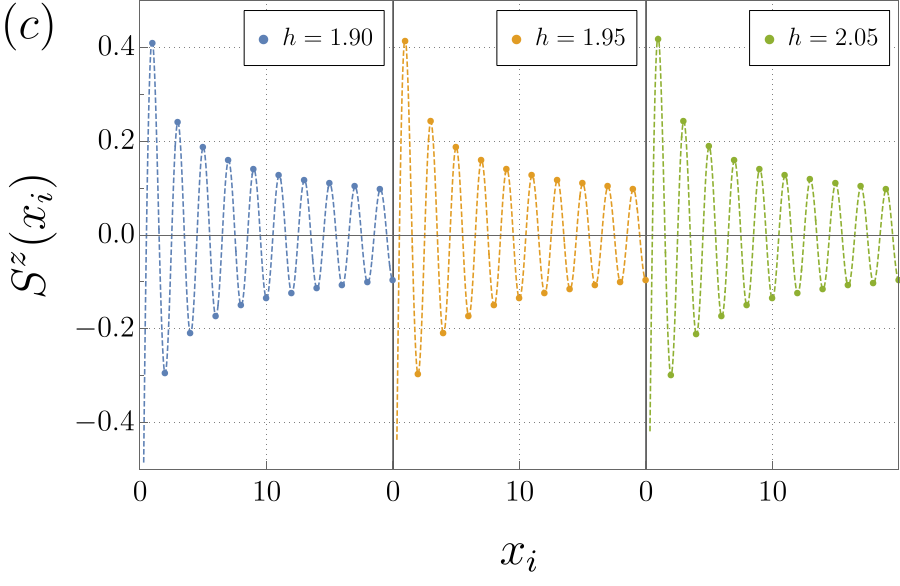}
    \caption{(a) The same figure from Fig.~\ref{fig:xi1}(b) together with the smooth interpolating function through the missing region. 
    (b) Fitting parameters $A$ and $B$ (see Eq.~\eqref{eq:fit}) with the substitution of $C$ and $D$ with the interpolated function obtained in (a). 
    (c) Magnetization data (dots) and its fit to Eq.~\eqref{eq:fit} (dashed lines) of the $h$ values of the shaded region using the parameters obtained in (a,b) for the first 20 sites of the $N = 1000$ chain. 
    The good agreement between the data and fit demonstrates the interpolation in (a) is valid.
    }
    \label{fig:fit}
\end{figure*}  

In Fig.~\ref{fig:xi1}(b) and Fig.~\ref{fig:xi2}(c) we have indicated the region where the fitting to Eq.~\eqref{eq:fit} numerically fails with a shade. 
Here, we show that this is merely due to numerical overfittings, and is not of physical consequence. 

First, we interpolate the data in Fig.~\ref{fig:xi1}(b) through the missing region. 
Some data points in the figure near $h_c = 2$ are additionally excluded in the interpolation to make the interpolated function smooth. 
The results, together with the original Fig.~\ref{fig:xi1}(b) are plotted in Fig.~\ref{fig:fit}(a). 
Now we take the values from the interpolated function for $C$ and $D$, and re-fit to Eq.~\eqref{eq:fit} to obtain $A$ and $B$. This result is shown in Fig.~\ref{fig:fit}(b) and in Fig.~\ref{fig:xi2}(b) in the main text. 
For the special point $h=2$, $C=D=0$ is used in the fit, and note that the resulting $A$ and $B$ fit parameters are smooth near $h=2$. 
The $A$, $B$ data away from $h=2$ in Fig.~\ref{fig:fit}(b) is identical to that obtained from the calculation in Fig.~\ref{fig:xi1}(b) and Fig.~\ref{fig:xi2}(c).

Finally, we use the newly fitted parameters $A$ to $D$ from Fig.~\ref{fig:fit}(a,b) and compare with the original $S^z(x_i)$ data for the shaded region, which are $h = 1.90,~1.95,~2.05$, in Fig.~\ref{fig:fit}(c). ($h=2.00$ data is included in Fig.~\ref{fig:xi1}(a)) 
The agreement between the data and fit is excellent in all three values of $h$. 
This demonstrates that the fitting failure at the shaded region is because of overfitting, and the true physics in those parameters connect smoothly to the behavior outside the region.
We also claim that small deviations from the interpolated function and data (for example, $C$ near $h=1.8$) are also the result of such numerical issues of overfit, while less serious than that more closer to $h=2$.

\vspace{.1in}

\end{widetext}
\end{document}